\documentclass[aps,prd,twocolumn,eqsecnum,amssymb,amsmath,showpacs,a4paper,superscriptaddress,nofootinbib,longbibliography]{revtex4-2}
\usepackage{graphicx}
\usepackage{amsfonts}
\usepackage{amsmath}
\usepackage{hyperref}
\usepackage{color}
\usepackage{dcolumn} 

\newcommand{\del}[2]{\frac{\partial #1}{\partial #2}}
\newcommand{\ud}{\mathrm{d}}

\begin{document}

\title{Energy Extraction From Non-Coalescing Black Hole Binaries}

\author{Lucas Timotheo Sanches}
 \email{lucas.t@ufabc.edu.br}
 \affiliation{Centro de Ci\^encias Naturais e Humanas, Universidade Federal do ABC (UFABC), 09210-170 Santo Andr\'e, S\~ao Paulo, Brazil}
\author{Maur\'icio Richartz}
 \email{mauricio.richartz@ufabc.edu.br}
\affiliation{Centro de Matem\'atica, Computa\c c\~ao e Cogni\c c\~ao, Universidade Federal do ABC (UFABC), 09210-170 Santo Andr\'e, S\~ao Paulo, Brazil}

\begin{abstract}
    We define and sketch the generalized ergosphere of the Majumdar-Papapetrou (MP) spacetime. In particular, we demonstrate the existence of closed orbits of negative energy that live outside the event horizon of such a spacetime. Relying on the Penrose process mechanism, we use these orbits to illustrate the possibility of energy extraction from a MP binary black hole by particle scattering. We also analyze the efficiency of the process, and construct explicit examples that optimize the extraction of energy. Lastly, we show how such concepts can be extended to a pair of non-coalescing Kerr black holes described by the Cabrera-Munguia, Manko and Ruiz (CMMR) metric.

\end{abstract}

\maketitle

\section {Introduction}

The best mathematical description of rotating black holes is given by the theory of General Relativity and, in particular, by the Kerr metric~\cite{Visser:2007fj,Bambi:2011mj,Teukolsky:2014vca,berti}. Unlike static black holes, a Kerr black hole is characterized by the existence of a very peculiar region around its event horizon, known as the \emph{ergosphere}. Particles that reach the ergosphere can still avoid the event horizon and, hence, are not doomed to end at the spacetime singularity. Nevertheless, any observer lying inside the ergosphere is unavoidably dragged along by the rotational motion of the black hole. 

Particles inside the ergosphere may have negative energies (according to a static observer at infinity). Relying on this property, Penrose and Floyd devised a mechanism (known as the Penrose process) to extract energy from a rotating black hole~\cite{PENROSE1971}. The idea consists in sending a particle from infinity towards the black hole and assumes that, once inside the ergosphere, it decays into two other particles. If one of the fragments is counter-rotating with the black hole and has negative energy, it will be captured by the black hole, meaning that the other fragment will escape to infinity. Due to the conservation of the four-momentum, the escaping fragment will have more energy and more angular momentum than the incident particle. Rotational energy and angular momentum are, thus, effectively extracted from the black hole.

To illustrate the importance of the Penrose mechanism, we highlight a few of the many recent research endeavours that attempt to establish a relation with observable astrophysical phenomena.
The collisional version of the process, for instance, considers multiple particles that collide and scatter in the ergoregion, allowing  arbitrarily high center-of-mass energies. This process can potentially act as a mechanism to eliminate dark matter particles near a supermassive black hole~\cite{Schnittman:2018ccg}. The magnetic Penrose process~\cite{Wagh1985,Tursunov:2019oiq}, on the other hand, considers charged particles and black holes surrounded by magnetic fields (originated, for instance, by plasma accretion disks around the black hole). The electromagnetic interaction allows for highly efficient energy extraction schemes, such as the one introduced in Ref.~\cite{Tursunov:2020juz} to model the emission of ultra-high energy cosmic rays from supermassive black holes.
Furthermore, recent numerical simulations of plasma and jets around Kerr black holes indicate the important role that negative energy particles and the Penrose process play on the total energy flux coming from the black hole's jets~\cite{Parfrey:2018dnc}. 

 The aim of our work is to investigate energy extraction via the Penrose process in a black hole binary system. Any study involving astrophysical binary systems must acknowledge the fact that there is no known \emph{exact} analytical solution of Einstein's equations describing such a system. Even if numerical or approximate analytical solutions (obtained, for example, through Numerical Relativity or Post Newtonian methods) are employed, the fact that the associated metric is non-stationary limits the applicability of the standard concept of an ergosphere. Recognizing these difficulties, in this paper we investigate the possibility of energy extraction from the Majumdar-Papapetrou (MP) metric~\cite{MAJUMDAR1947,PAPAPETROU1947}, which is an exact solution of Einstein's equations that describes a static binary of extremally charged black holes.
 
Despite its simplicity, the MP solution has recently been used as a surrogate model for black hole binaries in order to understand how single black hole phenomena transpose to a binary system. For instance, Ref.~\cite{ASSUMPCAO2018} has employed the MP metric to understand the connection between quasinormal modes and light rings in the context of black hole binaries. Refs.~\cite{Shipley:2016omi,Shipley:2019kfq}, on the other hand, have computed the shadows cast by a MP binary to better understand chaotic scattering in a binary black hole system. The resulting shadow shares many similarities and qualitative features with the shadows computed in Ref.~\cite{Bohn:2014xxa} using a numerically simulated binary black hole background. Ref.~\cite{BINI2019} has also applied the MP metric to analyze particle scattering around a black hole binary and asserted its effectiveness in approximating a head on collision in the limit of large separations and small approach speeds. We follow here an analog approach in order to gain physical intuition and qualitative insights about energy extraction from black hole binaries by the Penrose mechanism.  
In particular, we extend the concept of a particle dependent \emph{generalized ergosphere}~\cite{RUFFINI1971}, which enables the extraction of electromagnetic energy from Reissner-N\"ordstrom (RN) black holes, to the MP solution and study how the energy extraction efficiency is affected by the presence of a companion black hole.

Taking into account the fact that, in astrophysical contexts, any excess of electric charge in a black hole tends to be quickly neutralized~\cite{gibbons1975}, we also consider rotating systems in our work. More specifically, in order to illustrate how the main results for the MP spacetime can be extrapolated to a binary system composed of Kerr black holes, we employ an exact analytic solution of Einstein's field equations, discovered independently by Cabrera-Munguia, Manko and Ruiz (hereby referred to as the CMMR metric)~\cite{cabrera_metric,manko_ruiz_metric, manko_ruiz_thermo}. The CMMR metric describes two generic Kerr black holes that do not coalesce thanks to the presence of a ``strut'' that holds them apart at a fixed distance. In particular, we sketch the ergosphere of the CMMR spacetime for a selected set of parameters and give an example of a Penrose process around a binary of rotating black holes.


Our paper is organized as follows. In Sec.~II\,A we introduce the MP spacetime and derive the equations of motion for a charged massive particle. In Sec.~II\,B the notion of a generalized ergosphere is laid down and explored, while the existence of negative energy orbits is studied in Sec.~II\,C. In Sec.~III we show how a Penrose process can be set up in the MP spacetime and we give explicit examples of such a process. In Sec.~IV we analyze how much energy can be extracted and how it depends on the several parameters describing the system. Finally, in Sec.~V we illustrate how the concepts studied in the MP spacetime can be extended to the CMMR metric. The last section is dedicated to our final remarks.


\section{The Majumdar-Papapetrou spacetime} 

The MP spacetime is a static electrovacuum solution of Einstein's equations that represents a set of extremal black holes whose mutual gravitational attraction is cancelled by their mutual electromagnetic repulsion~\cite{MAJUMDAR1947,PAPAPETROU1947,HARTLE1972}.
 For two black holes of masses $M_1$ and $M_2$ and electric charges $Q_1=M_1$ and $Q_2=M_2$, in equilibrium and separated by a distance $2a$ along the $z$-axis, the MP line element, in Weyl's cylindrical coordinates $(t,\rho,\phi,z)$, is given by~\cite{SMERAK2016} 
\begin{equation}
  d s^2 = - \frac{d t^2}{U(\rho,z)^{2}} + U(\rho,z)^2\left[d\rho^2 + \rho^2 d\phi^2 + d z^2\right],
  \label{eq:majumdar_papapetrou_line_element}
\end{equation}
where 
\begin{equation}
  U(\rho,z) = 1 + \frac{M_1}{\sqrt{\rho^2 + (z+a)^2}} + \frac{M_2}{\sqrt{\rho^2 + (z-a)^2}}.
  \label{eq:mp_metric_potential_cylindric}
\end{equation}
 The electromagnetic potential $A_\mu$ associated with the MP solution is 
\begin{equation}
  A_\mu = \left(1 - \frac{1}{U}\right) \delta_{\mu t},
  \label{eq:electromagnetic_potential_mp}
\end{equation}
where $\delta_{\mu \nu}$ is the Kronecker delta. We note that Weyl's coordinates describe only the exterior of the black holes. The event horizons and the black holes themselves are collapsed into the points $\rho=0, \, z=\pm a$. We denote the total mass of the binary system by $M_T=M_1+M_2$ and its mass ratio by $M_R=M_2/M_1$. Without loss of generality, we assume that $0 \le M_R \le 1$. 

Being described by a static metric, the MP spacetime does not possess an ergosphere (in the usual sense of a spacetime region outside the black hole where the Killing vector $(\partial_t)^\mu$ becomes spacelike).
Consequently, the energy associated with geodesic motion in the MP spacetime is always positive, meaning that energy extraction by free particles is impossible. Charged particles, however, interact with charged black holes through Lorentz forces and, hence, do not follow geodesics. 
If the electromagnetic interaction is attractive, negative energy trajectories and energy extraction are, in principle, possible. For a single charged black hole described by the RN metric, the fact that the Penrose process is viable was shown in Ref.~\cite{DENARDO1973}. Building on the notion of a generalized ergosphere~\cite{RUFFINI1971,DENARDO1973}, we shall study the motion of charged particles around the MP spacetime to investigate the possibility of negative energy motion and energy extraction.

\subsection{Motion of charged particles in MP}
\label{sec:motion}

The motion of a massive charged test particle (with charge-to-mass ratio $\mu$) in a spacetime with metric $g_{\mu \nu}$, subject to the electromagnetic potential $A_{\mu}$, is determined by the Lagrangian 
\begin{equation}
  \mathcal{L} = \frac{1}{2}g_{\mu \nu}\dot{x}^\mu \dot{x}^\nu - \mu A_{\alpha}\dot{x}^\alpha,
  \label{eq:lagrangian_for_charged_particle}
\end{equation}
where $x^{\mu} = x^{\mu}(\lambda)$ denotes the position of the particle at proper time $\lambda$ and the dots represent derivatives with respect to $\lambda$. Adopting Weyl's cylindrical coordinates, and taking into account the explicit form of the MP metric, the Lagrangian above can be recast as~\cite{RYZNER2015}
\begin{equation}
  \mathcal{L} = \frac{1}{2}\left[-\frac{\dot{t}^2}{U^2} + U^2\left( \dot{\rho}^2 + \rho^2\dot{\phi}^2 + \dot{z}^2 \right) \right] - \mu\left(1-\frac{1}{U}\right)\dot{t}.
  \label{eq:explicit_lagrangian_for_charged_particle}
\end{equation}

Since the Lagrangian does not depend explicitly on $t$ and $\phi$, two constants of motion (as measured by freely-falling observers at infinity) can be readily identified. The constant associated with the time symmetry is the energy divided by the mass $m$ of the particle:
\begin{equation}
  E = -\del{\mathcal{L}}{\dot{t}} = \frac{\dot{t}}{U^2} + \mu\left(1-\frac{1}{U}\right).
  \label{eq:conserved_energy}
\end{equation}
The constant associated with the angular symmetry is the angular momentum per unit mass (with respect to the $z$ axis):
\begin{equation}
   L = \del{\mathcal{L}}{\dot{\phi}} = U^2 \rho^2 \dot{\phi}.
  \label{eq:conserved_momentum}
\end{equation}

Plugging the constants of motion into the 4-velocity normalization condition, i.e.~$\dot{x}^{\mu}\dot{x}_{\mu} = -1$, and solving for $E$, yields
\begin{equation}
  E = \mu\left(1-\frac{1}{U}\right) + \sqrt{\frac{L^2}{\rho^2U^4} + \frac{1}{U^2} + \dot{\rho}^2 + \dot{z}^2},
  \label{eq:alternative_expression_for_energy}
\end{equation}
where the negative root has been ignored due to the fact that $E$ must be positive for particles at infinity when $\mu = 0$ (see Ref.~\cite{RUFFINI1971} for a detailed discussion about positive and negative root states for $E$). We note that Eq.~\eqref{eq:alternative_expression_for_energy} reduces to the expression found in Ref.~\cite{DENARDO1973} for a RN black hole if the mass of one of the black holes is taken to be zero and an appropriate coordinate system, centered around the other black hole, is adopted. For convenience and later use, we rewrite Eq.~\eqref{eq:alternative_expression_for_energy} as
\begin{equation} \label{eq:effective1}
\dot{\rho}^2 + \dot{z}^2 = E_{\mathrm {eff}}^2(\rho,z) - V_{\mathrm {eff}}(\rho,z),
\end{equation}
where 
\begin{equation} \label{eq:effective2}
E_{\mathrm {eff}}(\rho,z) = E - \mu\left(1-\frac{1}{U}\right),  
\end{equation}
and
\begin{equation} \label{eq:effective3}
V_{\mathrm {eff}}(\rho,z) =\frac{L^2}{\rho^2U^4} + \frac{1}{U^2}.
\end{equation}
Note that the quantities defined above are subject to the following constraints: 
\begin{equation} \label{eq:effective_constraints}
E_{\mathrm {eff}}(\rho,z) \ge 0, \qquad E_{\mathrm {eff}}^2(\rho,z) \ge V_{\mathrm {eff}}(\rho,z).
\end{equation}

The equations of motion are obtained directly from the Euler-Lagrange equations. After using Eqs.~\eqref{eq:conserved_energy} and \eqref{eq:conserved_momentum} to eliminate $\dot{t}$ and $\dot{\phi}$, one is left with
\begin{align}
  \ddot{\rho} &-\frac{L^2(U +\rho \partial_{\rho}U)}{\rho^3 U^5} + \frac{2\dot{\rho}\dot{z}\partial_{z}U - (E^2 + \dot{z}^2 - \dot{\rho}^2)\partial_{\rho}U}{U} \nonumber \\ & \hspace{1.5cm} - \frac{\mu}{U}\left(\mu-2 E+\frac{E-\mu}{U}\right)\partial_{\rho}U = 0,
  \label{eq:ode_for_rho_motion}
\end{align}
and
\begin{align}
  \ddot{z} - \frac{L^2\partial_{z}U}{\rho^2 U^5} &+ \frac{2\dot{\rho}\dot{z}\partial_{\rho}U - (E^2 - \dot{z}^2 + \dot{\rho}^2)\partial_{z}U}{U} \nonumber \\ &-  \frac{\mu}{U}\left(\mu-2 E+\frac{ E-\mu}{U}\right)\partial_{z}U = 0,
  \label{eq:ode_for_z_motion}
\end{align}
which reduce to the equations of motion found in Ref.~\cite{ASSUMPCAO2018} for neutral particles. Together with Eqs.~\eqref{eq:conserved_energy} and \eqref{eq:conserved_momentum}, the two equations above fully determine the trajectory of a massive charged particle in the MP spacetime once appropriate initial conditions have been specified. In fact, we first specify the energy $E$, the angular momentum $L$, and the initial values for $\rho$, $z$, and $\dot z$. The initial value for $\dot{\rho}$ is determined from Eq.~\eqref{eq:alternative_expression_for_energy}. Using this information, we can solve the system of equations \eqref{eq:ode_for_rho_motion}-\eqref{eq:ode_for_z_motion} to obtain $\rho(\lambda)$ and $z(\lambda)$. The final step is the integration of Eqs.~\eqref{eq:conserved_energy} and \eqref{eq:conserved_momentum}, subject to the initial data for $t$ and $\phi$, to find $t(\lambda)$ and $\phi(\lambda)$.


\subsection{Generalized Ergosphere} \label{Sec:gen_ergo}

We want to know if charged particles can have negative energies in the MP spacetime. From Eq.~\eqref{eq:alternative_expression_for_energy} it is evident that, for fixed $\mu$, the energy is completely determined by the angular momentum $L$, and the values of $\rho$, $z$, $\dot{\rho}$ and $\dot{z}$  at any given instant of time. At a fixed position, the minimum possible energy is associated with particles at rest. Letting $\dot{\rho}=\dot{z}=0$ and $L=0$ (i.e.~$\dot{\phi}=0$), we conclude that the minimum energy (per unit mass) allowed for a particle sitting at $(t,\rho,\phi,z)$ is
 \begin{equation} \label{eq:minimum_energy}
  E_{\mathrm{min}} = \mu\left(1-\frac{1}{U}\right) + \frac{1}{U}.
 \end{equation}    
 
 This minimum energy will be negative if the following two conditions are satisfied. Since $U \ge 1$, $\mu$ must be negative, meaning that the charge of the particle must be opposite to the charge of the black holes. Additionally, the resting particle must be inside the spatial region determined by
\begin{equation} \label{eq:rescaled_negative_energy_condition}
 \frac{1}{\sqrt{\overline{\rho}^2 + (\overline{z}+1)^2}} + \frac{M_R}{\sqrt{\overline{\rho}^2 + (\overline{z}-1)^2}} > - \frac{1+M_R}{\overline{\mu}},
\end{equation}
where $\overline{\rho} = \rho/a$, $\overline{z} = z/a$ and $\overline \mu =  \mu M_T/a$ are dimensionless quantities. The parameter $\overline \mu$ can be understood as a measure of the potential energy (per unit particle mass) associated with the electromagnetic interaction between the particle and the binary.
The inequality above determines the generalized ergosphere of the MP spacetime: particles with opposite charge in relation to the black hole and located inside the region determined by Eq.~\eqref{eq:rescaled_negative_energy_condition} will have negative energies if their velocities are sufficiently small. Particles outside, on the other hand, will have positive energies regardless of their velocities.   

\begin{figure}[!htbp]
\centering
  \includegraphics[width = 0.95 \linewidth]{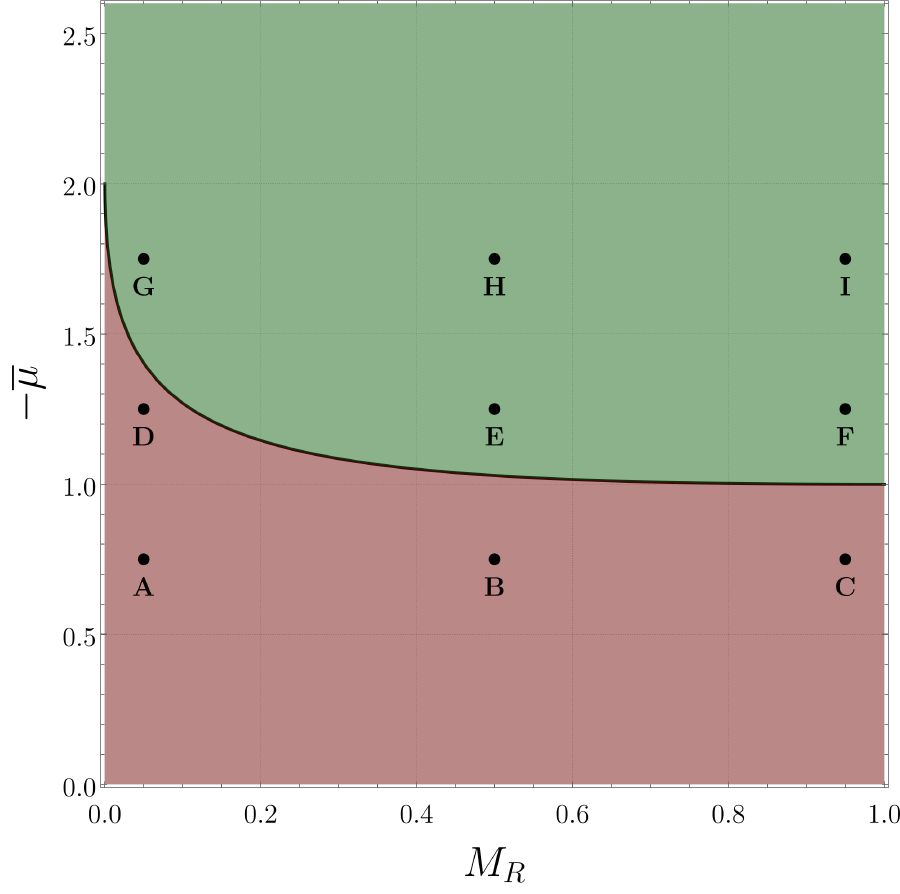}
  \caption{The parameter space $M_R$-$\overline \mu$. The generalized ergosphere exists if and only if $\overline \mu < 0$. It consists of a single connected region for any point inside the red section. For points in the green section, on the other hand, the ergosphere is the disjoint union of two connected regions. The black curve marks the boundary between connected and disconnected ergospheres. The ergospheres associated with the black labeled dots are plotted in Fig.~\ref{fig:ergos}.}
  \label{fig:regions} 
\end{figure}

\begin{figure}[!htbp]
 \centering
  \includegraphics[width = 0.95 \linewidth]{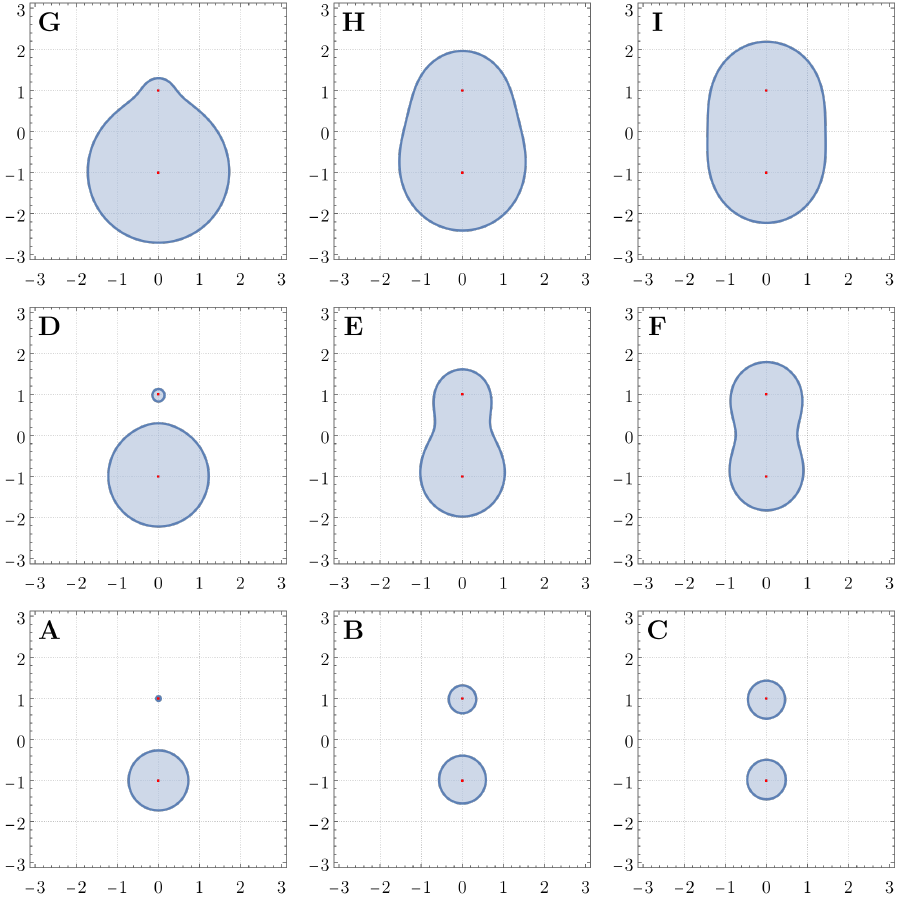}
  \caption{The $\phi=0$ section of the ergosphere for different points (labeled \textbf{A}-\textbf{I}) in the parameter space of Fig.~\ref{fig:regions}. In each plot the horizontal and vertical axes are $\overline \rho$ and $\overline z$, respectively. The red dots indicate the location of the black holes.}
  \label{fig:ergos}
\end{figure}

 Note that the notion of a generalized ergosphere depends not only on the geometry of the spacetime, but also on properties of the particle (through the charge-to-mass ratio $\mu$), as in Refs.~\cite{RUFFINI1971,DENARDO1973}. In particular, the shape of the ergosphere depends only on the parameters $M_R$ and $\overline \mu$. In order to understand this dependence, we plot the generalized ergosphere for nine pairs $M_R$-$\overline \mu$. Each pair corresponds to a point in the parameter space of Fig.~\ref{fig:regions} and is labeled by a letter (\textbf{A}-\textbf{I}). The $\phi=0$ section of the associated ergospheres
are shown in Fig.~\ref{fig:ergos}. Due to the axisymmetry of the problem, the ergospheres are the solids of revolution obtained by the rotation of the regions shown in Fig.~\ref{fig:ergos} with respect to the $\overline z$-axis. Note that the generalized ergosphere can be either a single connected region (for parameters inside the red section of Fig.~\ref{fig:regions}) or the disjoint union of two connected regions\footnote{In such a case we shall refer to each connected region as a single ergosphere.} (for parameters in the green section of Fig.~\ref{fig:regions}). The boundary separating connected and disconnected ergospheres, represented by the black line in Fig.~\ref{fig:regions}, corresponds to the saturation of Eq~\eqref{eq:rescaled_negative_energy_condition}.

What information can we extract from Eq.~\eqref{eq:rescaled_negative_energy_condition} and Figs.~\ref{fig:regions} and \ref{fig:ergos}? Naturally, the ergosphere becomes more evenly distributed around the black holes when the mass ratio $M_R$ approaches one. This is seen in the figures by following the points \textbf{A} $\rightarrow$ \textbf{B} $\rightarrow$ \textbf{C}, or \textbf{D} $\rightarrow$ \textbf{E} $\rightarrow$ \textbf{F}, or \textbf{G} $\rightarrow$ \textbf{H} $\rightarrow$ \textbf{I}. Furthermore, if $-2 < \overline \mu < -1$, we see that an increase in the mass ratio may produce a single ergosphere from two disjoint ones, as in \textbf{D} $\rightarrow$ \textbf{E}. Similarly, for any mass ratio, an increase in the absolute value of $\overline \mu$ can also induce the merger of the ergospheres, as in \textbf{D} $\rightarrow$ \textbf{G}, \textbf{B} $\rightarrow$ \textbf{E} and \textbf{C} $\rightarrow$ \textbf{F}. In fact, no matter what the mass ratio is, if $-\overline \mu$ is sufficiently small, each black hole will be surrounded by its own ergosphere. In contrast, if $-\overline \mu$ is sufficiently large, there will be a single ergosphere surrounding the binary black hole. Finally, we remark that, according to Eq.~\eqref{eq:rescaled_negative_energy_condition}, the effects of the charge of the particle, of the total mass of the binary and of the separation parameter are all combined in $\overline \mu$. Therefore, as far as the visualization of the ergosphere is concerned, the effect of increasing $|\mu|$ is exactly the same as increasing $M_T$ or decreasing $a$.

\begin{figure}[!htbp]
  \centering
  \includegraphics[width = 0.98 \linewidth]{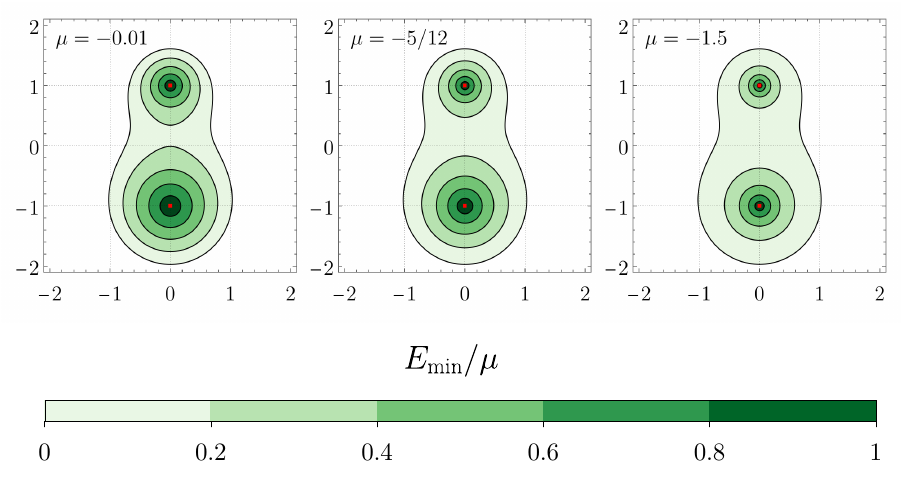}
  \caption{Energy levels inside the generalized ergosphere of a MP black hole with $M_R=1/2$ and $\overline \mu = -5/4$, corresponding to the point \textbf{E} in Figs.~\ref{fig:regions} and \ref{fig:ergos}. The color bar represents $E_{\mathrm min}/\mu$. From left to right, the panels correspond respectively to $\mu=-0.01$, $\mu=-5/12$, and $\mu=-1.5$. The horizontal and vertical axes in each panel are $\overline \rho$ and $\overline z$, respectively. The red dots indicate the location of the black holes.} 
\label{fig:energylevels} 
\end{figure}

The minimum energy (per unit mass) at a given point of the spacetime, in contrast, depends on the explicit values of $\mu$, $M_R$ and $M_T/a$. To illustrate this dependence, we plot in Fig.~\ref{fig:energylevels} the energy levels inside the ergosphere determined by $M_R=1/2$ and $\overline \mu = -5/4$ (which correponds to the point \textbf{E} in Figs.~\ref{fig:regions} and \ref{fig:ergos}) for different combinations of $\mu$ and $M_T/a$. The color bar in Fig.~\ref{fig:energylevels} indicates the value of $E_{\mathrm{min}}/\mu$, which, according to Eq.~\eqref{eq:minimum_energy}, varies from zero (at the boundary of the ergosphere) to one (at the black holes). As $|\mu|$ decreases and $M_T/a$ increases (while $\overline \mu$ and $M_R$ are kept fixed), the shape of the ergosphere remains unchanged, but the energy levels become more evenly distributed around the black holes.  

\subsection{Negative energy trajectories}
\label{sec:neg_energy}
Trajectories associated with negative energies are confined within the ergosphere defined by Eq.~\eqref{eq:rescaled_negative_energy_condition}. For a generic set of initial conditions, such paths will eventually end up at either one of the black holes. This is not much different than the case of a single RN or a Kerr black hole, in which particles with negative energies that start outside the event horizon necessarily enter the black hole and reach the spacetime singularity. Instead of focusing on this type of trajectory, in this section we concentrate on trajectories which have no analogs in standard (Kerr and RN) black hole spacetimes~\cite{Grib:2013hxa,Zaslavskii:2020crn}. We want to investigate whether or not closed orbits of negative energy that live outside the black holes are allowed\footnote{We note, however, that if the Kerr black hole is immersed in an external magnetic field, closed orbits of negative energy are allowed outside the event horizon~\cite{PRASANA1982,FELICE2004}. We also note that stable circular orbits of negative energy exist around a Kerr naked singularity (and that the Penrose process can effectively take place using such orbits)~\cite{STUCHLIK1980}.}. 
 We follow the procedure described in the last paragraph of section \ref{sec:motion} to solve the equations of motion. To simplify our work, we restrict our attention to two classes of planar motion and show that, by fine tuning the initial data, one is able to find closed orbits of negative energies in the MP spacetime. These closed orbits of negative energies are unstable, in the sense that generic perturbations of the initial conditions result in trajectories that end at one of the black holes. 

\begin{figure}[!htbp]
\centering
   \includegraphics[width = 0.95 \linewidth]{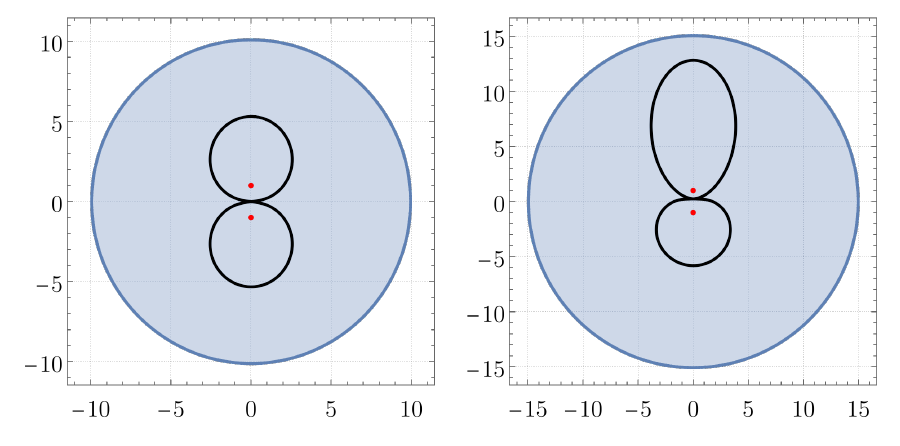}
   \caption{Example of closed trajectories of negative energy (black curves) in the plane containing the black holes. The blue region is the generalized ergosphere of the spacetime, while the red dots indicate the location of the black holes. The horizontal and vertical axes in each panel are $\overline \rho$ and $\overline z$, respectively. Left planel: trajectory corresponding to $L=0$, $E=-2/10$, $a=M_1=M_2=1$, $\mu = -5$, $\rho(0) = 0$, $z(0) = 5.314064237978...$ (fine tuned), and $\dot{z}(0) = 0$. The associated period of revolution is $\lambda \approx 22.64$.  Right planel: trajectory corresponding to $L=0$, $E=-2/100$, $a=1$, $M_1=2 M_2= 2$, $\mu = -5$, $\rho(0) = 0$, $z(0) = 12.832155724084...$ (fine tuned), and $\dot{z}(0) = 0$. The associated period of revolution is $\lambda \approx 43.28$. }
    \label{fig:orbit_closed_bh}
    \end{figure}

The first class of orbits assumes zero angular momentum ($L=0$) so that the trajectories are restricted to a plane that contains both black holes (here we choose the $\phi=0$ plane without loss of generality). For a given set of parameters $a$, $M_1$, $M_2$, $E$, and $\mu$, after fixing $\rho(0) = 0$ and $\dot{z}(0) = 0$, we fine tune the initial position $z(0)$. The initial value $\dot{\rho} (0)$ is determined from Eq.~\eqref{eq:alternative_expression_for_energy}. To illustrate, we show in Fig.~\ref{fig:orbit_closed_bh} two examples of eight-shaped orbits obtained with this scheme. The left panel of Fig.~\ref{fig:orbit_closed_bh} represents an orbit with energy (per unit mass) $E=-2/10$ and charge-to-mass ratio $\mu=-5$ for an equal mass MP binary ($M_1=M_2=1$) with separation parameter $a=1$. The trajectory starts on the $z$-axis, at $z(0) = 5.314064237978...$ (fine tuned). One complete revolution of the trajectory, with corresponding period of $\lambda \approx 22.64$, is shown. The right panel of Fig.~\ref{fig:orbit_closed_bh}, on the other hand, exhibits an orbit of $E=-2/10$ and $\mu=-5$ for a MP binary with mass ratio $M_R=1/2$ ($M_1=2,M_2=1$) and separation parameter $a=1$. The trajectory starts on the $z$-axis, at $z(0) = 12.832155724084...$ (fine tuned). One complete revolution of the trajectory, with correponding period of approximately $\lambda \approx 43.28$, is shown.

The second class of orbits is only possible for equal mass binaries and comprises trajectories that are confined to the $z=0$ plane, which is the plane equidistant to the black holes. Particles with negative energy whose motion is constrained to this plane will evolve in a perpetual oscillatory motion, either in straight lines (when $L=0$) or in more complicated precessing orbits (when $L \neq 0$). Even though these trajectories are unstable to generic perturbations, on the plane itself they are stable. In other words, infinitesimal perturbations of $\rho(0)$ and $\dot \rho(0)$ produce infinitesimal variations on the original trajectory (if $z(0)$ and $\dot z(0)$ are kept fixed).

\begin{figure}[!htbp]
   \centering
   \includegraphics[width =  0.95 \linewidth]{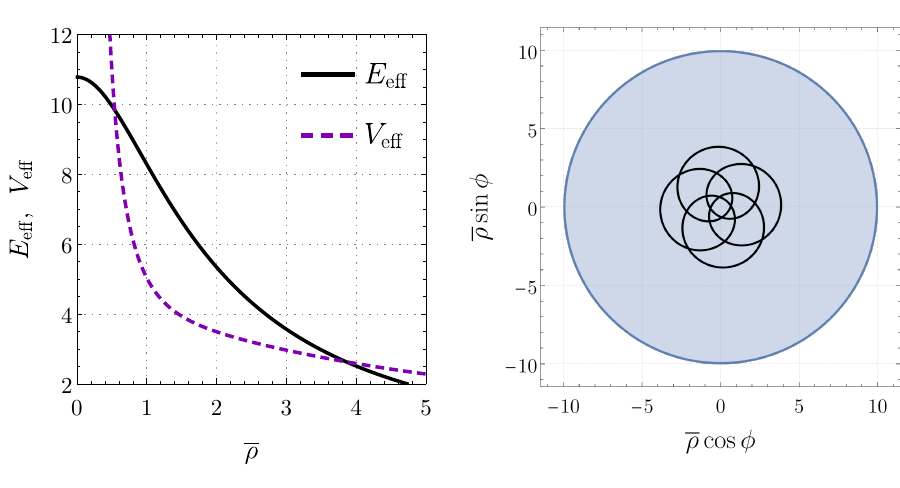}
   \caption{Left panel: effective energy (black curve) and effective potential (purple dashed curve) for $L=12.85869$, $E=-5/100$, $a=1$, $M_1 = M_2 = 1$, $\mu = -5$. Right panel: an example of a closed trajectory of negative energy in the $z = 0$ symmetry plane. The blue region is the generalized ergosphere of the spacetime. The trajectory (black curve) is generated by setting the parameters to $L=12.85869$, $E=-5/100$, $a=1$, $M_1 = M_2 = 1$, $\mu = -5$, $\rho(0) = 2$, and $\phi(0) = 0$. The associated period of revolution is $\lambda \approx 35$.}
    \label{fig:orbit_closed_z}
    \end{figure}

To analyze these orbits we resort to Eqs.~\eqref{eq:effective1}-\eqref{eq:effective3}.
Given $a$, $E$ and $L$, the critical points $\rho$ where $E_{\mathrm{eff}}(\rho,0)^2 = V_{\mathrm{eff}}(\rho,0)$ represent circular closed orbits. Typically, however, the particle will be confined inside a compact section of the $z=0$ plane, in a characteristic ``zoom-whirl'' orbit~\cite{ASSUMPCAO2018,Levin:2008mq}, moving between a minimum radius $\rho_{\mathrm min}$ and a maximum radius $\rho_{\mathrm max}$. In Fig.~\ref{fig:orbit_closed_z}, we show the effective energy and the effective potential for a MP spacetime with $a=1$ and $M_1 = M_2 = 1$, when the particle's charge-to-mass ratio, angular momentum (per unit mass), and energy (per unit mass) are, respectively, $\mu=-5$, $L=12.85869$, and $E=-5/100$. The particle is constrained to move between $\rho_{\mathrm min}\approx 0.543$ and $\rho_{\mathrm max}\approx 3.857$. The corresponding trajectory, assuming the starting point to be $\rho(0)=2$ and $\phi(0)=0$, is also shown in Fig.~\ref{fig:orbit_closed_z}. 

\section{Penrose Process}

Following Penrose's original proposal~\cite{PENROSE1971} and its extension to RN black holes~\cite{RUFFINI1971,DENARDO1973}, we now investigate the possibility of energy extraction from a MP binary black hole. The mechanism we shall explore is analogous to the most standard one: a negatively charged particle is sent towards the binary black hole and, once inside the generalized ergosphere, breaks up into two fragments, one of which escapes to infinity with more energy than the original particle. We assume that the incident particle follows a trajectory $T_{(0)}$, which starts outside the ergosphere and ends inside it, at the break-up point $(\rho_*,\phi_*,z_*)$. From the break-up point, two other trajectories, labeled $T_{(1)}$ and $T_{(2)}$, start. Each trajectory $T_{(i)}$ is a timelike path $x^{\mu}_{(i)}(\lambda)$ parametrized by its proper time $\lambda$. To fix notation, let $m_{(i)}$, $\mu_{(i)}$, $E_{(i)}$, $L_{(i)}$, and $P^{\mu}_{(i)}$ denote, respectively, the mass, the charge-to-mass ratio, the energy per unit mass, the angular momentum per unit mass (with respect to the z-axis), and the 4-momentum of the particle $i$ on trajectory $T_{(i)}$. We assume that fragment $1$ remains inside the ergosphere (meaning that $E_{(1)} < 0 $), while fragment $2$ escapes back to infinity.

The quantities that characterize each particle are related by conservation equations. Charge conservation, for instance, yields 
\begin{equation}
  \mu_{(0)} m_{(0)}  = \mu_{(1)} m_{(1)} + \mu_{(2)} m_{(2)}.
  \label{eq:charge_conservation}
\end{equation}
The conservation of the four-momentum applied at the break-up point, on the other hand, reads
\begin{equation}
  P^{\mu}_{(0)} = P^{\mu}_{(1)} + P^{\mu}_{(2)}.
  \label{eq:four_momentum_conservation}
\end{equation}
Each component of the vector equation \eqref{eq:four_momentum_conservation} above corresponds to a different conservation equation with straightforward physical interpretation. The zero-component is just the conservation of total energy, i.e.
\begin{equation}
 E_{(0)}m_{(0)} = E_{(1)}m_{(1)} + E_{(2)}m_{(2)}.
  \label{eq:conservation_of_energy}
\end{equation}
The spatial components are the conservations of linear momenta, i.e. 
\begin{equation}
 \begin{cases}
  m_{(0)}\dot{\rho}_{(0)} = m_{(1)}\dot{\rho}_{(1)} +m_{(2)}\dot{\rho}_{(2)} \\
   m_{(0)}\dot{z}_{(0)} = m_{(1)}\dot{z}_{(1)} + m_{(2)}\dot{z}_{(2)} 
\end{cases},
  \label{eq:conservation_mom_lin}
\end{equation}
and the conservation of angular momentum, i.e.
\begin{equation}
  L_{(0)}m_{(0)} = L_{(1)} m_{(1)} + L_{(2)}m_{(2)}.
  \label{eq:conservation_mom_ang}
\end{equation}
We remark that all derivatives in \eqref{eq:conservation_mom_lin} are evaluated at the break-up point. We also note that the break-up of the incident particle naturally imposes restrictions on the masses of its fragments. By squaring Eq.~\eqref{eq:four_momentum_conservation} and using the fact that the four-momentum is future-pointing and timelike, we obtain the inequality~\cite{bhat1985energetics}
\begin{equation} \label{eq:mass_constraint}
m_{(1)}^2 + m_{(2)}^2 < m_{(0)}^2.
\end{equation} 

\begin{figure}[!htbp]
\centering
  \includegraphics[width = 0.95 \linewidth]{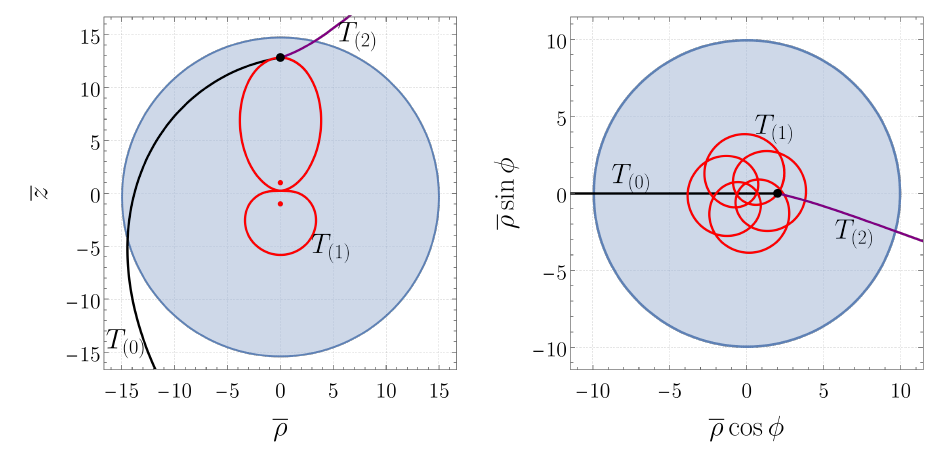}
  \caption{Left panel: Penrose process in the $\rho$-$z$ plane of a MP spacetime with $M_1=2$, $M_2=1$, and $a=1$. Right panel: Penrose process in the $z=0$ plane of a MP spacetime with $M_1=M_2=1$ and $a=1$. In each plot, the incoming trajectory $T_{(0)}$ (black curve) splits at the black point into the negative energy orbit $T_{(1)}$ (red curve) and the trajectory of the escaping fragment $T_{(2)}$ (purple curve). The parameters that generate these trajectories are shown in Tables \ref{tab:example1} and \ref{tab:example2}. The blue region is the generalized ergosphere (for particle 1) and the red dots indicate the location of the black holes.}
  \label{fig:example1} 
\end{figure}

\renewcommand{\arraystretch}{1.2}
\begin{table}[!htbp]
\begin{ruledtabular}
\begin{tabular}{lllllll}
$i$ & $m_{(i)}$ & $\mu_{(i)}$ & $E_{(i)}$  & $L_{(i)}$  & $\dot{\rho}_{(i)}$ & $\dot{z}_{(i)}$ \\ \vspace{-0.3cm} \\
0 & 1.00000 & -0.08345 & 1.00000 & 0 & 0.60000 & 0.09699 \\
1 & 0.70000 & -5.00000 & -0.02000 & 0 & 0.41207 & 0.00000  \\
2 & 0.23248 & 14.69629 & 4.36171 & 0 & 1.34012 & 0.41719  \\
\end{tabular}
\end{ruledtabular}
\caption{Parameters that generate the trajectories $T_{(0)}$, $T_{(1)}$, and $T_{(2)}$ of the Penrose process shown in the left panel of Fig.~\ref{fig:example1}. The derivatives $\dot{\rho}_{(i)}$ and $\dot{z}_{(i)}$ are evaluated at the break-up point.}
\label{tab:example1}
\end{table}

\renewcommand{\arraystretch}{1.2}
\begin{table}[!htbp]
\begin{ruledtabular}
\begin{tabular}{lllllll}
$i$ & $m_{(i)}$ & $\mu_{(i)}$ & $E_{(i)}$  & $L_{(i)}$  & $\dot{\rho}_{(i)}$ & $\dot{z}_{(i)}$ \\ \vspace{-0.3cm} \\
0 & 1.00000 & -0.27698 & 1.00000 & 0.00000 & 1.00000 & 0 \\
1 & 0.10000 & -5.00000 & -0.05000 & 12.85870 & 1.36059 & 0  \\
2 & 0.33342 & 0.66890 & 3.01423 & -3.85662 & 2.59116 & 0  \\
\end{tabular}
\end{ruledtabular}
\caption{Parameters that generate the trajectories $T_{(0)}$, $T_{(1)}$, and $T_{(2)}$ of the Penrose process shown in the right panel of Fig.~\ref{fig:example1}. The derivatives $\dot{\rho}_{(i)}$ and $\dot{z}_{(i)}$ are evaluated at the break-up point.}
\label{tab:example2}
\end{table}

From Eq.~\eqref{eq:conservation_of_energy}, the energy carried away by the escaping fragment is $E_{(2)}m_{(2)} = E_{(0)}m_{(0)} - E_{(1)}m_{(1)}$. Since we have assumed that $E_{(1)} < 0$, the escaping particle will carry away more energy than the incident particle had. If the negative energy fragment collapses to one of the black holes, it will directly decrease the energy associated with the black hole, as in Penrose's original proposal. If, on the other hand, the negative energy fragment remains in a closed orbit (such as the ones described in Sec.~\ref{sec:neg_energy}), where does the extra energy for the escaping fragment come from? Since the final state (binary black hole and bound particle fragment) is less energetic than the binary itself, we conclude that the extracted energy is associated with the binding energy of the binary. In other words, the presence of the bound particle fragment reduces the energy required to form a MP binary and the excess energy is transferred to the escaping fragment.

In Fig.~\ref{fig:example1} we exhibit examples of Penrose processes that include some of the negative energy orbits shown in Sec.~\ref{sec:neg_energy}. The left panel exhibits a Penrose process that takes place in the $\rho$-$z$ plane of a MP spacetime with $M_1=2$, $M_2=1$, and $a=1$. The incoming particle breaks-up at $\overline \rho=0$ and $\overline z = 5.314064237978...$ (fine tuned). The right panel, on the other hand,  exhibits a Penrose process that occurs in the $z=0$ plane of a MP spacetime with $M_1=M_2=1$ and $a=1$. The incoming particle breaks-up at $\overline \rho=2$ and $\phi = 0$. The parameters that generate these examples and satisfy Eqs.~\eqref{eq:charge_conservation}-\eqref{eq:mass_constraint} are given in Tables \ref{tab:example1} and \ref{tab:example2}.

\section{Energy extraction efficiency}

The efficiency $\eta$ of the Penrose process can be defined as the ratio between the energy output and the energy input. Using Eq.~\eqref{eq:conservation_of_energy}, we have
\begin{equation}
  \eta = \frac{E_{(2)}m_{(2)} - E_{(0)}m_{(0)}}{E_{(0)}m_{(0)}} = - \frac{E_{(1)}m_{(1)}}{E_{(0)}m_{(0)}}.
  \label{eq:penrose_efficiency_general}
\end{equation}
A natural question arises: what is the maximum efficiency of the Penrose process in a MP spacetime? Since $\eta$ is directly proportional to $E_{(1)}$ and inversely proportional to $E_{(0)}$, in order to maximize the efficiency of the process we need to make the absolute value of $E_{(1)}$ as large as possible and $E_{(0)}$ as small as possible. We also want the negative energy fragment to be as massive as possible in comparison to the mass of the incident particle. In other words, we want to extract as much energy as possible starting with as little energy as possible. We shall assume that the MP spacetime is fixed (meaning that $M_1$, $M_2$ and $a$ are known), the break-up point has been specified as $(\rho_*, z_*, \phi_*)$ and the charge-to-mass ratio $\mu_{(1)}$ is known. Given these hypotheses, we will determine how much energy can be extracted from a MP black hole and how the remaining parameters must be chosen in order to optimize the process.

\subsection{Maximum efficiency}

First of all, the minimum energy for an incident particle coming from infinity, according to Eq.~\eqref{eq:alternative_expression_for_energy}, is $E_{(0)}=1$ and corresponds to the particle having zero kinetic energy at infinity. For that reason, we shall assume from now on that $L_{(0)}=0$ and $E_{(0)}=1$. Secondly, according to Eq.~\eqref{eq:minimum_energy} and the discussion in the first paragraph of Sec.~\ref{Sec:gen_ergo}, the energy per unit mass of particle $1$ is most negative if the particle is initially at rest. Therefore, at the break-up point we set 
\begin{equation} \label{eq:traj1}
\dot{\rho}_{(1)}=\dot{z}_{(1)}=\dot{\phi}_{(1)}=0,
\end{equation}  
meaning that the associated angular momentum per unit mass and energy per unit mass are, respectively, $L_{(1)}=0$ and 
\begin{equation} \label{eq:penrose_neg_energy}
E_{(1)}=E_{(1)}^{\mathrm {min}}=\mu_{(1)}\left(1 - \frac{1}{U_*} \right) + \frac{1}{U_*},
\end{equation}
where we have defined $U_*=U(\rho_*,z_*)$ for simplicity.

Let us now study the allowed values for $m_{(1)}$. The conservation of linear momenta, Eq.~\eqref{eq:conservation_mom_lin}, yields the relation
\begin{equation} \label{eq:mass2_eq1}
m_{(2)}^2  = m_{(0)}^2 \left( \frac{\dot \rho_{(0)} ^2 +  \dot z_{(0)} ^2 }{\dot \rho_{(2)} ^2 + \dot z_{(2)} ^2 } \right) + m_{(1)}^2 \left( \frac{\dot \rho_{(1)} ^2 + \dot z_{(1)} ^2 }{\dot \rho_{(2)} ^2 + \dot z_{(2)} ^2 } \right),
\end{equation}
where all derivatives are evaluated at the break-up point.
After replacing $\dot \rho_{(i)} ^2 + \dot z_{(i)} ^2$ using Eqs.~\eqref{eq:effective1}-\eqref{eq:effective3}, and employing the conservation equations  \eqref{eq:charge_conservation}, \eqref{eq:conservation_of_energy} and \eqref{eq:conservation_mom_ang}, the expression above reduces to
\begin{equation} \label{eq:mass2_eq2}
m_{(2)} = \sqrt{m_{(0)}^2 - 2 m_{(0)} m_{(1)} \alpha_{(0)} + m_{(1)}^2},
\end{equation}
where 
\begin{equation} \label{eq:alpha_definition}
\alpha_{(0)} =  U_* \left[ 1 - \mu_{(0)}\left(1 - \frac{1}{U_*}\right) \right].
\end{equation}
Note that the expression between brackets in the definition of $\alpha_{(0)}$ is precisely the effective energy of the incident particle evaluated at the break-up point as given by Eq.~\eqref{eq:effective2}. The inequalities given in Eq.~\eqref{eq:effective_constraints}, together with the fact that $U_* \ge 1$, therefore imply that 
\begin{equation} \label{eq:alpha_constraint}
\alpha_{(0)} \ge 1,   
\end{equation}
and
\begin{equation} \label{eq:mu0max}
\mu_{(0)} \le 1.
\end{equation}

Eq.~\eqref{eq:mass2_eq2} and the fact that the masses are positive, together with the constraint imposed by Eq.~\eqref{eq:mass_constraint}, also imply that
\begin{equation}\label{eq:bound_of_m1_m0}
0 < \frac{m_{(1)}}{m_{(0)}} < \alpha_{(0)} . 
\end{equation}
Since the radicand in Eq.~\eqref{eq:mass2_eq2} must be positive, the bound given by Eq.~\eqref{eq:bound_of_m1_m0} can be further refined to yield
\begin{equation} \label{eq:m1_max}
0 < m_{(1)} < m_{(0)} \left( \alpha_{(0)} - \sqrt{\alpha_{(0)}^2 - 1} \right).
\end{equation}
Hence, the allowed range for $m_{(1)}$ is maximized when the inequalities \eqref{eq:alpha_constraint} and \eqref{eq:mu0max} are saturated. In fact, when $\mu_{(0)} \rightarrow 1$ one can choose $m_{(1)}\rightarrow m_{(0)}$, thus maximizing the ratio $m_{(1)}/m_{(0)}$ that appears in Eq.~\eqref{eq:penrose_efficiency_general}. Consequently, the efficiency of the Penrose process in a MP spacetime is bound from above according to
\begin{equation} \label{eq:eta_max_theory}
\eta < \eta ^{b}  = -E_{(1)}^{\mathrm {min}}.
\end{equation} 
We remark that the upper bound above is a function of $\mu_{(1)}$, the break-up point coordinates $\rho_*$ and $z_*$, and the MP parameters ($M_1$, $M_2$, and $a$). With the help of Eq.~\eqref{eq:penrose_neg_energy}, we now investigate in detail how these quantitities affect the efficiency bound.  

\subsection{Dependence on the parameters}

\begin{figure}[!htbp]
\centering
  \includegraphics[width = 0.95 \linewidth]{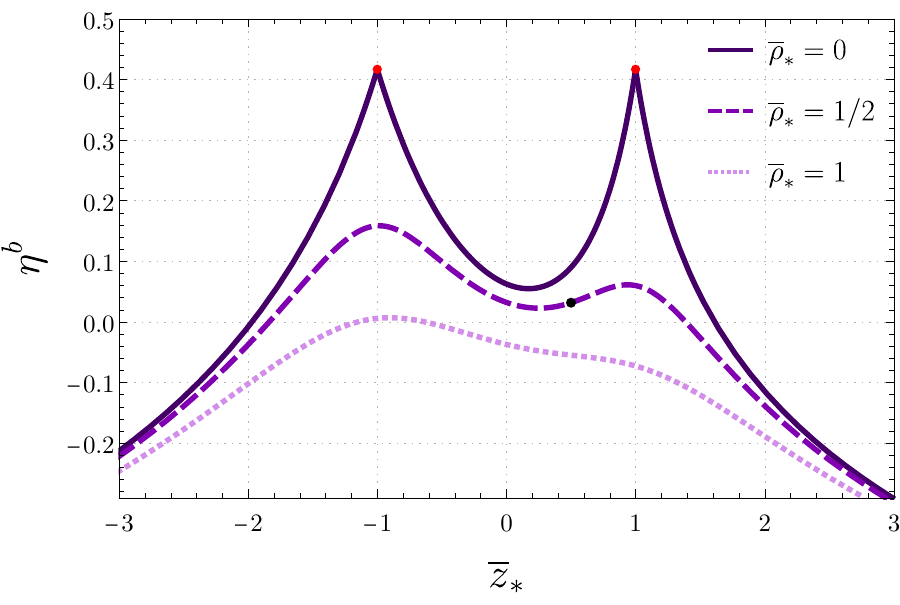}
  \caption{Efficiency bound $\eta ^ b$ as a function of $z_*$ for $\overline \rho_*=0$ (solid curve), $\overline \rho_*=1/2$ (dashed curve), and $\overline \rho_*=1$ (dotted curve) when $M_R=1/2$, $M_T=3$, and $\overline \mu_{(1)} = -5/4$. The associated ergosphere corresponds to the point \textbf{E} in Figs.~\ref{fig:regions} and \ref{fig:ergos}, while the associated energy levels are shown are shown in the middle panel of Fig.~\ref{fig:energylevels}. The red dots indicate the efficiency when the particle breaks-up at the black holes. The black dot shows the maximum efficiency $\eta^b = 0.03161$ for the processes represented in Fig.~\ref{fig:examplemax1} and Tables \ref{tab:max1} and \ref{tab:max2}.}
  \label{fig:efficiency0} 
\end{figure}

The dependence of $\eta^{b}$ on the charge-to-mass ratio $\mu_{(1)}$ is simple: $\eta^{b}$ increases linearly with $\mu_{(1)}$. The dependence of $\eta^{b}$ on the break-up point can be understood with the help of the energy levels shown in Fig.~\ref{fig:energylevels}: the efficiency bound increases as the break-up point approaches either one of the black holes. A more detailed analysis is shown in Fig.~\ref{fig:efficiency0}, where we plot $\eta^{b}$ as a function of $z_*$ for selected values of $\rho_*$ when $M_R=1/2$, $M_T=3$, and $\overline \mu_{(1)} = -5/4$ (corresponding to the energy levels shown in the middle panel of Fig.~\ref{fig:energylevels}). Note that when the break-up point is outside the ergosphere, the efficiency bound becomes negative, meaning that the escaping fragment will carry away less energy than the incident particle had. The Penrose process is most efficient if the break-up occurs exactly at either one of the event horizons, as indicated by the red dots in Fig.~\ref{fig:efficiency0}. When this happens, the upper bound is $-\mu_{(1)}$, which is in agreement with results obtained for the RN metric~\cite{bhat1985energetics,parthasarathy1986high}. 

\begin{figure}[!htbp]
\centering
  \includegraphics[width = 0.95 \linewidth]{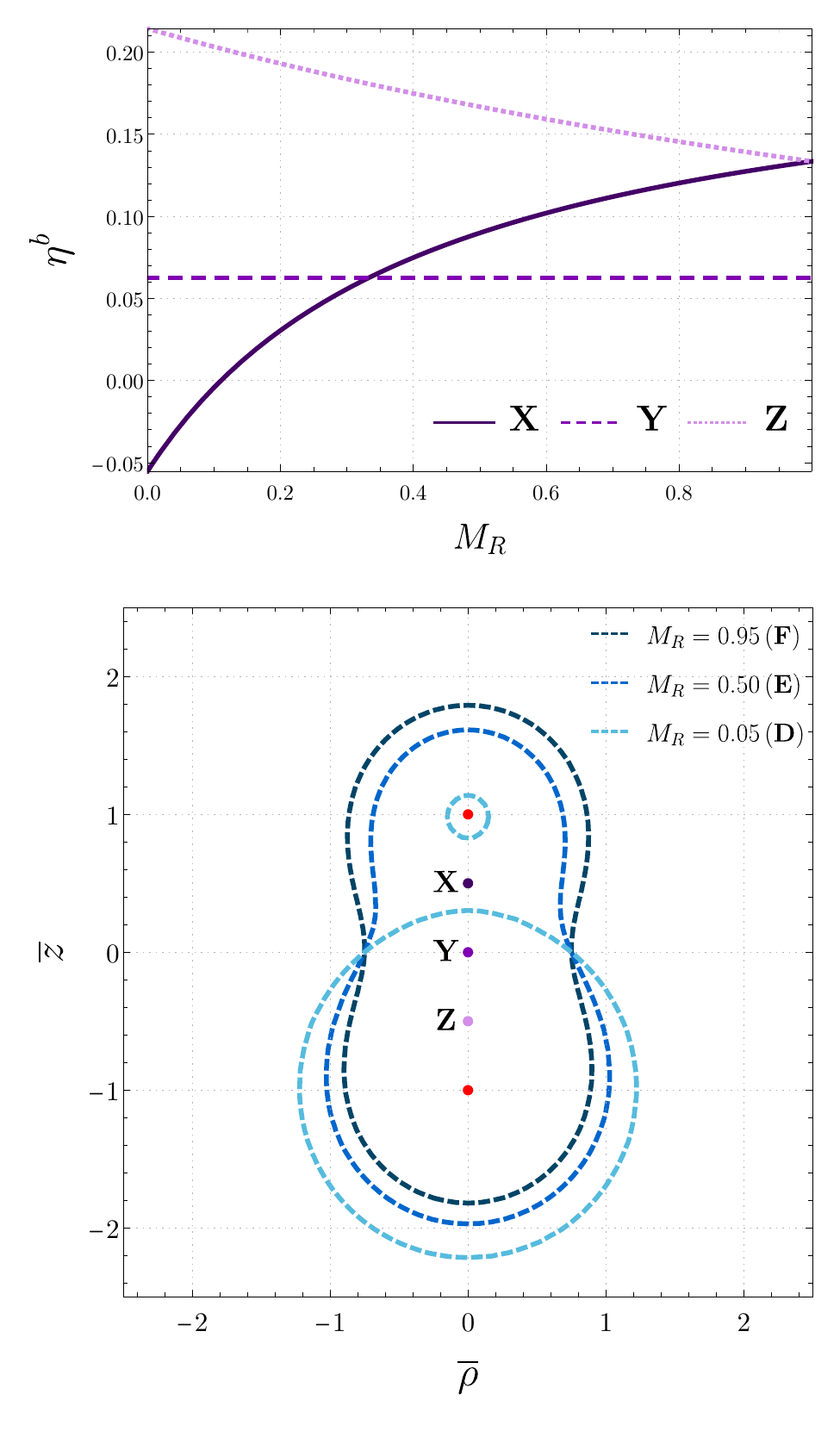}
  \caption{Top panel: efficiency bound $\eta^{b}$ as a function of $M_R$ when $M_T/a=3$ and $\mu_{(1)} = -5/12$. Bottom panel: generalized ergosphere for selected values of $M_R$ (corresponding to the points  \textbf{D}, \textbf{E} and \textbf{F} in Fig.~\ref{fig:regions}) when $M_T/a=3$ and $\mu_{(1)} = -5/12$. The red dots indicate the location of the black holes. The purple dots are the locations of the break-up points $(\overline \rho_*,\overline z_*)=(1/2,1/2)$, $(\overline \rho_*,\overline z_*)=(1/2,0)$, and $(\overline \rho_*,\overline z_*)=(1/2,-1/2)$ and are labeled \textbf{X}, \textbf{Y} and \textbf{Z}, respectively. Each curve in the top panel corresponds to one of the break-up points shown in the bottom panel.}
  \label{fig:efficiency2} 
\end{figure}

\begin{figure}[!htbp]
\centering
  \includegraphics[width = 0.95 \linewidth]{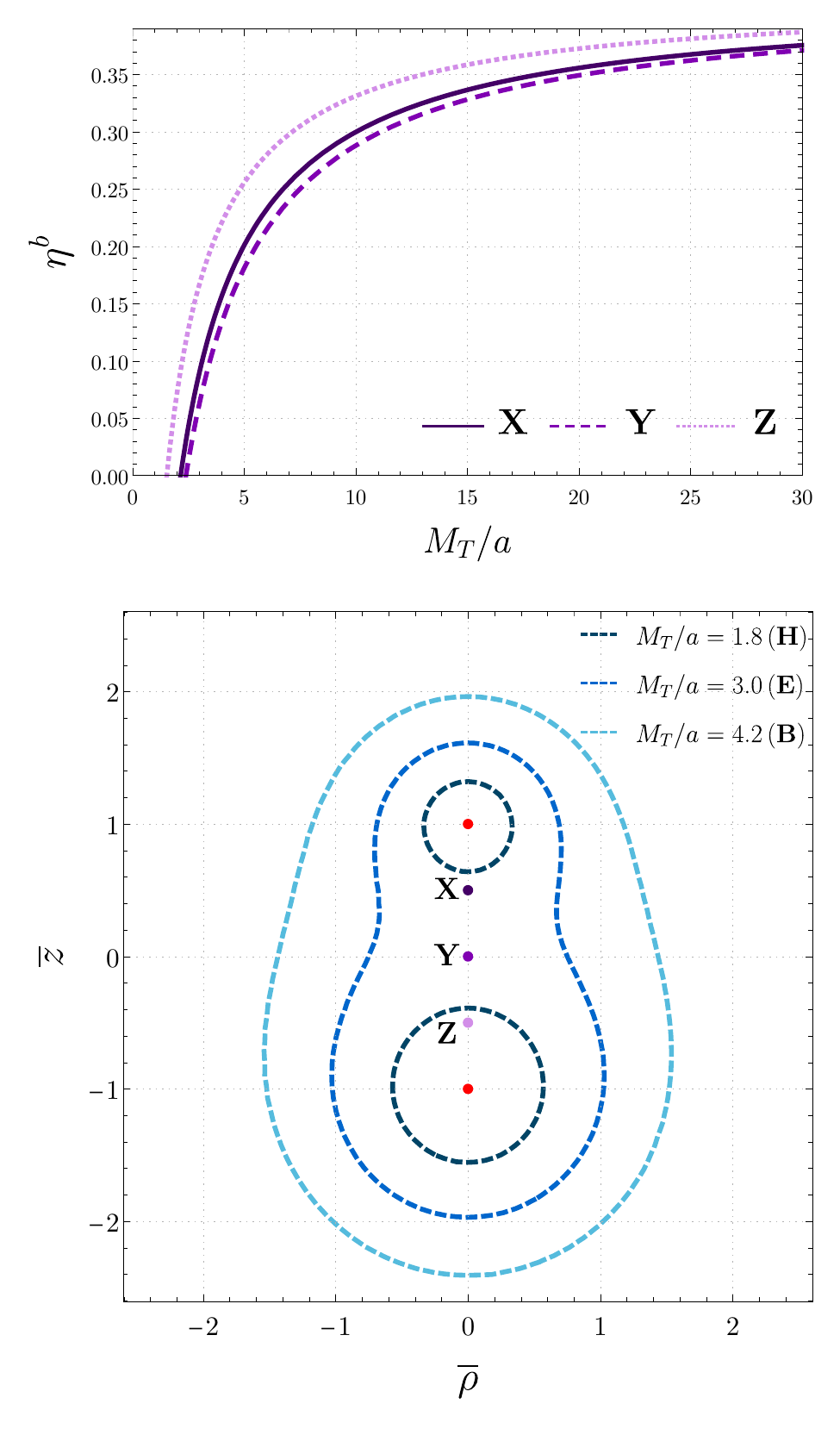}
  \caption{Top panel: efficiency bound $\eta^{b}$ as a function of $M_T/a$ when $M_R=1/2$ and $\mu_{(1)} = -5/12$. Bottom panel: generalized ergosphere for selected values of $M_T/a$ (corresponding to the points  \textbf{B}, \textbf{E} and \textbf{H} in Fig.~\ref{fig:regions}) when $M_R=1/2$ and $\mu_{(1)} = -5/12$. The red dots indicate the location of the black holes. The purple dots are the locations of the break-up points $(\overline \rho_*,\overline z_*)=(1/2,1/2)$, $(\overline \rho_*,\overline z_*)=(1/2,0)$, and $(\overline \rho_*,\overline z_*)=(1/2,-1/2)$ and are labeled \textbf{X}, \textbf{Y} and \textbf{Z}, respectively. Each curve in the top panel corresponds to one of the break-up points shown in the bottom panel.}
  \label{fig:efficiency3} 
\end{figure}

To study the dependence of the efficiency bound on the masses of the black holes, we plot $\eta ^{b}$ as a function of the mass ratio $M_R$ when $M_T/a=3$ (top panel of Fig.~\ref{fig:efficiency2}) and as a function of $M_T/a$ when $M_R=1/2$  (top panel of Fig.~\ref{fig:efficiency3}). In both cases we have set the charge-to-mass ratio as $\mu_{(1)} = -5/12$. Each curve in these plots is associated with a different break-up point. The break-up points are shown in the bottom panels of Figs.~\ref{fig:efficiency2} and \ref{fig:efficiency3}. In the bottom panel of Fig.~\ref{fig:efficiency2}, we also exhibit the generalized ergosphere associated with a few selected values of $M_R$ (which are chosen to reproduce the ergospheres labeled \textbf{D}, \textbf{E} and \textbf{F} in Figs.~\ref{fig:regions} and \ref{fig:ergos}). Analogously, the ergospheres depicted in the bottom panel of Fig.~\ref{fig:efficiency3} (and the associated values of $M_T/a$) correspond to the ergospheres labeled \textbf{B}, \textbf{E} and \textbf{H} in Figs.~\ref{fig:regions} and \ref{fig:ergos}. 

As shown in the top panel of Fig.~\ref{fig:efficiency2}, when the mass ratio $M_R$ increases (while the other parameters are kept fixed), $\eta ^{b}$ will also increase if the break-up point is closer to the lighter black hole and will decrease if the break-up point is closer to the heavier black hole. This behaviour is related to the fact that the growth of $M_R$ produces an expansion of the ergosphere around the lighter companion and a reduction of the ergosphere around the heavier companion, as seen in the bottom panel of Fig.~\ref{fig:efficiency2}.
 Note that the efficiency bound is independent of the mass ratio if the break-up point is equidistant to the black holes.
 On the other hand, as shown in Fig.~\ref{fig:efficiency3}, when $M_T/a$ increases (while the other parameters are kept fixed), the ergosphere expands and $\eta ^{b}$ increases. We note that in the limit $M_T/a \rightarrow \infty$, no matter where the break-up occurs, the efficiency bound approaches its maximum, i.e. $\eta ^{b} \rightarrow - \mu_{(1)}$. This is explained by the fact that the distance between the break-up point and the black holes becomes negligible in comparison to the size of the ergosphere when $M_T/a \rightarrow \infty$. 

\begin{figure}[!htbp]
\centering
  \includegraphics[width = 0.95 \linewidth]{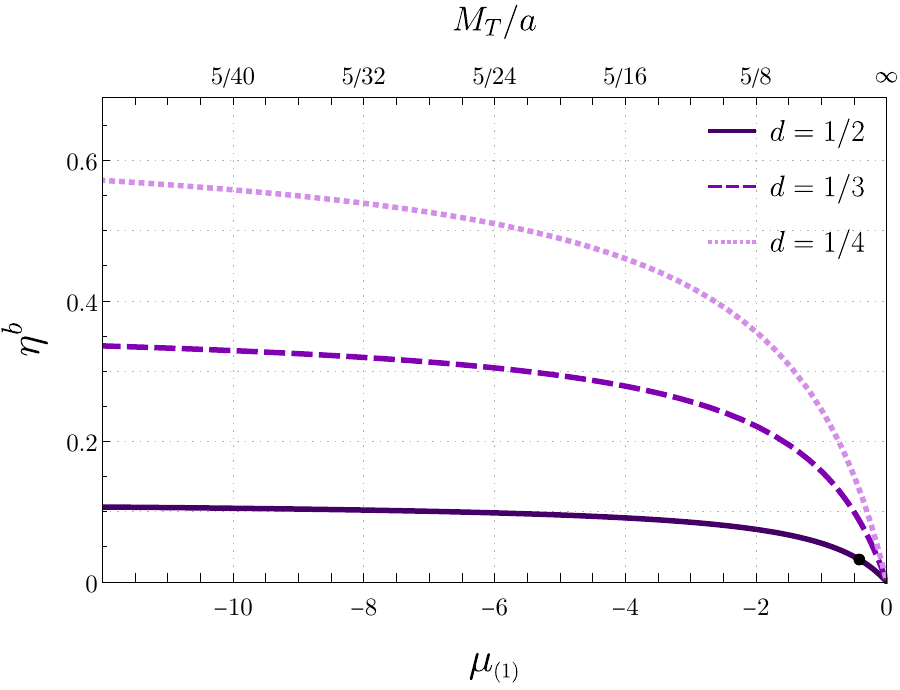}
  \caption{Efficiency bound $\eta ^ b$ as a function of $\mu_{(1)}$ (bottom axis) and $M_T/a$ (top axis) for $M_R=1/2$ and $\overline \mu _{(1)} = -5/4$. The fact that $\overline \mu _{(1)}$ is fixed implies that $\mu_{(1)}$ and $M_T/a$ are inversely proportional to each other. The curves, from bottom to top, correspond to the break-up points $(\overline \rho_*, \overline z_*)=(d,1-d)$, with $d=1/2$, $d=1/3$, and $d=1/4$ respectively. The black dot shows the maximum efficiency $\eta^b = 0.03161$ for the processes represented in Fig.~\ref{fig:examplemax1} and Tables \ref{tab:max1} and \ref{tab:max2}, i.e.~when $d=1/2$, $\mu_{(1)} = - 5/12$ and $M_T/a=3$.}
  \label{fig:efficiency1} 
\end{figure}

Finally, we investigate the behaviour of $\eta ^{b}$ when $\mu_{(1)}$ and $M_T/a$ vary simultaneously, but their product, i.e. $\overline \mu_{(1)} = \mu_{(1)} M_T/a$, is kept fixed, meaning that the shape of the ergosphere is fixed (only the energy levels inside of it change). In Fig.~\ref{fig:efficiency1} we plot $\eta^b$ as a function of $\mu_{(1)}$ and $M_T/a$ for three different break-up points when $M_R=1/2$ and $\overline \mu_{(1)} = -5/4$, so that the resulting ergosphere corresponds to the one labeled \textbf{E} in Figs.~\ref{fig:regions} and \ref{fig:ergos}. The break-up points are chosen to be $(\overline \rho_*,\overline z_*)=(d,1-d)$, with $d=1/2$, $d=1/3$, and $d=1/4$. We observe that, as $|\overline {\mu}|$ increases and $M_T/a$ decreases, the efficiency bound increases, approaching an asymptotic value in the limit $\mu_{(1)} \rightarrow -\infty$ and $M_T/a \rightarrow 0$.

\subsection{Examples}

We now give concrete examples of energy extraction in a MP binary black hole spacetime whose efficiency approaches the theoretical maximum given by Eq.~\eqref{eq:eta_max_theory}. The general procedure that we follow is outlined below. First, we choose the MP parameters $M_1$, $M_2$ and $a$. Second, we choose the charge-to-mass ratio $\mu_{(1)}$ and a break-up point $(\rho_*,z_*,\phi_*)$ that is inside the generalized ergosphere of the spacetime. Without loss of generality, we set $m_{(0)}=1$.

Following the discussion leading to the inequality \eqref{eq:eta_max_theory}, we set $L_{(0)}=L_{(1)}=L_{(2)}=0$, meaning that all trajectories are restricted to the plane $\phi=\phi_*$. The energy $E_{(1)}$ is determined by Eq.~\eqref{eq:penrose_neg_energy}, while $E_{(0)}=1$. According to Eqs.~\eqref{eq:mu0max} and \eqref{eq:m1_max}, we choose 
\begin{eqnarray}
\label{eq:epsilondef} \mu_{(0)} &= 1 - \varepsilon, \\ 
\label{eq:nudef} m_{(1)} &= 1 - \nu,
\end{eqnarray}
where $\varepsilon$ and $\nu$ are small positive parameters satisfying
\begin{equation} \label{eq:nuepsilonrelation}
\nu > \varepsilon (U_* - 1) \left( \sqrt{1 + \frac{2}{\varepsilon (U_* - 1)}} - 1 \right).
\end{equation}      
Since $m_{(1)}$  has been fixed, the mass $m_{(2)}$ can be determined by Eq.~\eqref{eq:mass2_eq2}. The quantities $\mu_{(2)}$ and $E_{(2)}$ are determined from Eqs.~\eqref{eq:charge_conservation} and \eqref{eq:conservation_of_energy}, respectively.

At the break-up point, according to Eq.~\eqref{eq:traj1}, we have $\dot \rho_{(1)}=\dot z_{(1)} = 0$ for the negative energy fragment. For the incident particle, on the other hand, the choices for $\mu_{(0)}$ and $E_{(0)}$ imply that   
\begin{equation}\label{eq:velocity_eq_max_energy_example}
\dot \rho_{(0)} ^2 + \dot z_{(0)} ^2 = \varepsilon \frac{U_*-1}{U_* ^2} \left[2 + \varepsilon (U_* - 1) \right].
\end{equation}
By choosing the angle $\theta_{(0)} = \mathrm{Arg}\left(\dot \rho_{(0)} + i \, \dot z_{(0)} \right)$ between the velocities $\dot \rho_{(0)}$ and $\dot z_{(0)}$, equation Eq.~\eqref{eq:velocity_eq_max_energy_example} can be used to determine $\dot \rho_{(0)}$ and $\dot z_{(0)}$ individually at the break-up point. The conservation of linear momentum then fixes the values of $\dot \rho_{(2)}$ and $\dot z_{(2)}$ through Eq.~\eqref{eq:conservation_mom_lin}. At this point, the trajectories $T_{(0)}$, $T_{(1)}$, and $T_{(2)}$ have all been determined and the efficiency of the associated Penrose process is $\eta=(1-\nu)\eta^{b}$.  In order to maximize the efficiency of the Penrose process, one must, therefore, set $\eta$ to be as small as possible. Note, however, that one cannot choose $\varepsilon=0$, because this would imply $\dot \rho_{(0)} = \dot z_{(0)} = 0$ and  $\ddot \rho_{(0)} = \ddot z_{(0)} = 0$ at the break-up point, contradicting the fact the trajectory $T_{(0)}$ starts infinitely far away. Similarly, $\nu$ cannot be chosen to saturate inequality \eqref{eq:nuepsilonrelation}, otherwise $m_{(2)}$ would be exactly zero, contradicting the fact that all trajectories are timelike. If desired, this can be remedied by assuming, from the start, that the escaping fragment is massless (however, this would modify Eq.~\eqref{eq:alternative_expression_for_energy} and the present analysis). Nevertheless, it is possible, in principle, to have infinitesimally small values for $\varepsilon$ and $\nu$. Finally, we point out that not every angle $\theta_0$ produces trajectories that are consistent with the assumptions of a Penrose process. More precisely, only certain ranges of $\theta_0$ give rise to trajectories  $T_{(0)}$ and $T_{(2)}$ that, respectively, start and end infinitely far away from the black holes.

\begin{figure}[!htbp]
\centering
  \includegraphics[width = 0.95 \linewidth]{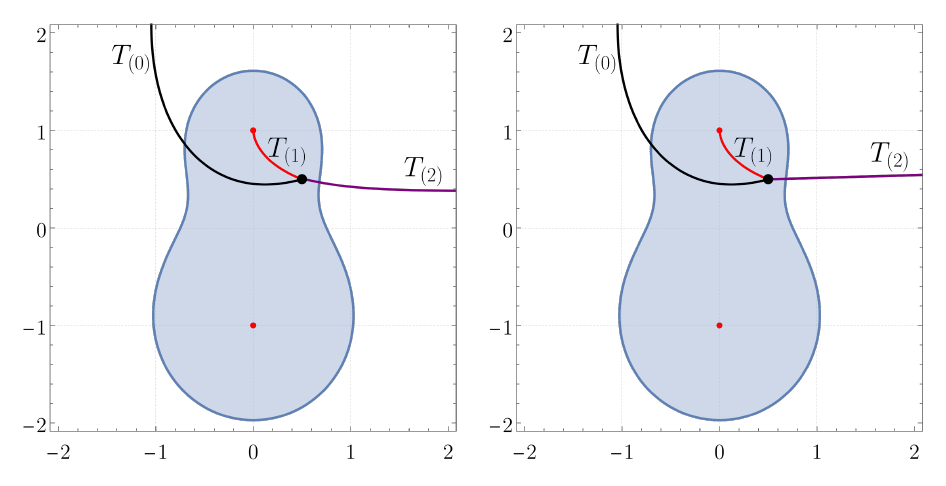}
  \caption{Examples of Penrose processes that approach the maximum theoretical efficiency in a MP spacetime. The efficiencies of the processes in the left and right panel are, respectively, $90 \%$ and $99 \%$ of the theoretical maximum. In both cases the incoming trajectory $T_{(0)}$ (black curve) splits at the black point into the negative energy orbit $T_{(1)}$ (red curve) and the trajectory of the escaping fragment $T_{(2)}$ (purple curve). The parameters that generate these trajectories are shown in Tables \ref{tab:max1} and \ref{tab:max2}. The blue region is the generalized ergosphere (for particle 1) and the red dots indicate the location of the black holes.}
  \label{fig:examplemax1} 
\end{figure}

\renewcommand{\arraystretch}{1.2}
\begin{table}[!htbp]
\begin{ruledtabular}
\begin{tabular}{lllllll}
$i$ & $m_{(i)}$ & $\mu_{(i)}$ & $E_{(i)}$  & $L_{(i)}$  & $\dot{\rho}_{(i)}$ & $\dot{z}_{(i)}$ \\ \vspace{-0.3cm} \\
0 & 1.00000 & 0.99990 & 1.00000 & 0 & 0.00609 & 0.00158 \\
1 & 0.90000 & -0.41667 & -0.03161 & 0 & 0.00000 & 0.00000  \\
2 & 0.09756 & 14.09301 & 10.54183 & 0 & 0.06243 & 0.01621  \\
\end{tabular}
\end{ruledtabular}
\caption{Parameters that generate the trajectories $T_{(0)}$, $T_{(1)}$, and $T_{(2)}$ of the Penrose process shown in the left panel of Fig.~\ref{fig:examplemax1}. The derivatives $\dot{\rho}_{(i)}$ and $\dot{z}_{(i)}$ are evaluated at the break-up point.
}
\label{tab:max1}
\end{table}

\renewcommand{\arraystretch}{1.2}
\begin{table}[!htbp]
\begin{ruledtabular}
\begin{tabular}{lllllll}
$i$ & $m_{(i)}$ & $\mu_{(i)}$ & $E_{(i)}$  & $L_{(i)}$  & $\dot{\rho}_{(i)}$ & $\dot{z}_{(i)}$ \\ \vspace{-0.3cm} \\
0 & 1 & 0.99999 & 1.00000 & 0 & 0.00193 & 0.00050 \\
1 & 0.99000 & -0.41667 & -0.03161 & 0 & 0.00000 & 0.00000  \\
2 & 0.00685 & 206.13521 & 150.50460 & 0 & 0.28104 & 0.07297  \\
\end{tabular}
\end{ruledtabular}
\caption{Parameters that generate the trajectories $T_{(0)}$, $T_{(1)}$, and $T_{(2)}$ of the Penrose process shown in the right panel of Fig.~\ref{fig:examplemax1}. The derivatives $\dot{\rho}_{(i)}$ and $\dot{z}_{(i)}$ are evaluated at the break-up point.
}
\label{tab:max2}
\end{table}

We conclude by showing in Fig.~\ref{fig:examplemax1} two explicit examples of the procedure outlined above for the MP spacetime with $M_1=2$, $M_2=1$, and $a=1$. The charge-to-mass ratio of the negative energy fragment is chosen as $\mu_{(1)}=-5/12$, so that the associated generalized ergosphere is the one identified by the letter \textbf{E} in Figs.~\ref{fig:regions} and \ref{fig:ergos}. The  break-up point is chosen as $(\rho_*,z_*)=(1/2,1/2)$, fixing the trajectory $T_{(1)}$ and its energy (per unit mass) $E_{(1)}=-0.03161$ (remember that the initial conditions are chosen so that $E_{(1)}$ is the minimum possible). According to Eq.~\eqref{eq:eta_max_theory}, the efficiency bound is $\eta^{b} = 0.03161$ (which corresponds to the black dots in Figs.~\ref{fig:efficiency0} and \ref{fig:efficiency1}). The trajectories $T_{(0)}$ and $T_{(2)}$ are specified by the choices of $\varepsilon$, $\nu$, and $\theta_{(0)}$. In the left panel of Fig.~\ref{fig:examplemax1}  we exhibit the Penrose process for $\varepsilon=10^{-4}$, $\nu=10^{-1}$, and $\theta_{(0)}=0.25403$, whose efficiency is $90 \%$ of the theoretical maximum. In the right panel of Fig.~\ref{fig:examplemax1} we exhibit the Penrose process for $\varepsilon=10^{-5}$, $\nu=10^{-2}$, and $\theta_{(0)}=0.25403$, whose efficiency is $99 \%$ of the theoretical maximum.  The parameters that generate these examples and satisfy Eqs.~\eqref{eq:charge_conservation}-\eqref{eq:mass_constraint} are  given in Tables \ref{tab:max1} and \ref{tab:max2}.  


\section{Kerr binaries}

We now consider the extension of our previous results to a binary system of rotating black holes described by the CMMR metric. In Weyl's cylindrical coordinates, the CMMR line element reads
\begin{align}
  \ud s^2 &= -f(\rho,z)\left[\ud t - \omega(\rho,z)\ud\phi\right]^2 \nonumber \\
  & + f(\rho,z)^{-1}\left[e^{2\gamma(\rho,z)}\left(\ud \rho^2 + \ud z^2\right) + \rho^2\ud\phi^2\right],
  \label{eq:cmmr_line_element}
\end{align}
where the real valued functions $f(\rho,z)$, $\omega(\rho,z)$ and $\exp[2\gamma(\rho,z)]$ are defined as in Sec.~IV of Ref.~\cite{manko_ruiz_thermo}. As in the case of the MP metric, only the exterior of the black holes is described by these coordinates. In particular, the outer event horizons of the constituent black holes are straight lines in these coordinates (see Fig.~\ref{fig:cmmr_ergospheres}).

The CMMR solution is fully characterized by five independent parameters, namely the masses $M_{1,2}$, the angular momenta per unit mass $a_{1,2}$ and the coordinate distance $R$ between the black hole centers. From these, we define three additional parameters, namely $M_T = M_1 + M_2$, which represents the total mass of the system, $J_T = M_1 a_1 + M_2 a_2$, which represents the total angular momentum of the system, and $a_*$, which is a root of the cubic equation
\begin{align}
    \left(R^2 - M_T^2 + a_*^2\right)&\left(a_1 + a_2 - a_*\right) \nonumber  \\  + \, & 2 \left(R + M_T \right)\left(J_T - M_T a_* \right) = 0.
    \label{eq:cubic_eq_for_a}
\end{align}
%

We note that, depending on the parameters, the CMMR metric can represent a black hole-black hole binary, a naked singularity-naked singularity binary, or a black hole-naked singularity binary~\cite{cabrera_metric,manko_ruiz_metric, manko_ruiz_thermo}. In our analysis, the chosen parameters always correspond to a binary black hole solution. In practice, this means that the chosen parameters must: (i) produce real valued and positive horizon lengths\footnote{The horizon half-lengths are given by the expressions $\sigma_1$ and $\sigma_2$ in Sec.~IV of Ref.~\cite{manko_ruiz_thermo}.}, (ii) produce horizons that do not touch or overlap, and (iii) produce a single real root for $a_*$ in Eq.~\eqref{eq:cubic_eq_for_a}. 


\subsection{Geodesics}

Following the procedure on Sec.~IIA, we will make use of the Lagrangian formalism to determine the geodesic equations and the conserved quantities corresponding to the symmetries of the system. The Lagrangian associated with the geodesic motion of a massive and neutral test particle in the CMMR metric is~\cite{Dubeibe2016}
\begin{equation}
  2\mathcal{L} = -f(\dot{t} - \omega\dot{\phi})^2 +f^{-1}\left[e^{2\gamma}\left( \dot{\rho}^2 + \dot{z}^2 \right) + \rho^2\dot{\phi}^2 \right],
  \label{eq:cmmr_lagrangian}
\end{equation}
where, once again, dots represent derivatives with respect to the proper time $\lambda$.

Due to the stationarity and the axisymmetry of the system, we can identify two constants of the motion analogous to the quantities defined in Eqs.~\eqref{eq:conserved_energy}  and \eqref{eq:conserved_momentum}. The energy per unit mass, as measured by a static observer at infinity, is given by
\begin{equation}
  E = f\dot{t} - \omega f\dot{\phi},
  \label{eq:cmmr_energy_t_phi}
\end{equation}
and the angular momentum (with respect to the $z$ axis) per unit mass, as measured by a static observer at infinity, is given by
\begin{equation}
  L = \omega f \dot{t} + \left( \frac{\rho^2}{f} - \omega^2 f \right)\dot{\phi}.
  \label{eq:cmmr_ang_mom_t_phi}
\end{equation}

Using Eqs.~\eqref{eq:cmmr_energy_t_phi} and \eqref{eq:cmmr_ang_mom_t_phi} to eliminate $\dot{t}$ and $\dot{\phi}$ from the the normalization of the four velocity, i.e. $\dot{x}^\mu\dot{x}_\mu = -1$, we obtain an expression for the energy $E$ in  terms of $\dot{\rho}$, $\dot{z}$ and the angular momentum $L$:
\begin{align}
  E  = & \frac{-f^2\omega L}{\rho^2-\omega^2 f^2}  + \left[ \frac{\rho^2e^{2\gamma}(\dot{\rho}^2 + \dot{z}^2)}{\rho^2 - \omega^2f^2}  \right. \nonumber \\
  &  \left. + \left( \frac{\rho fL}{\rho^2 - \omega^2f^2} \right)^2 + \frac{\rho^2f}{\rho^2 - \omega^2f^2} \right]^{1/2},
  \label{eq:cmmr_energy_rho_z}
\end{align}
where the positive sign is once again chosen for the square root in order to guarantee that a static particle at infinity has positive energy. We rewrite the equation above as Eq.~\eqref{eq:effective1}, where the effective energy and the effective potential are now given by
\begin{equation}
  V_{\text{eff}} = \frac{\rho^2 - \omega^2 f^2}{\rho^2 e^{2\gamma}}\left[ \left( \frac{\rho fL}{\rho^2 - \omega^2f^2} \right)^2 + \frac{\rho^2f}{\rho^2 - \omega^2f^2}\right]
  \label{eq:cmmr_effective_potential}
\end{equation}
and
\begin{equation}
  E_{\text{eff}} = \frac{\rho^2 - \omega^2 f^2}{\rho^2 e^{2\gamma}}\left(E + \frac{f^2\omega L}{\rho^2-\omega^2 f^2}\right)^2.
  \label{eq:cmmr_effective_energy}
\end{equation}
Note that the constraints given in Eq.~\eqref{eq:effective_constraints} also apply to Eqs.~\eqref{eq:cmmr_effective_potential} and \eqref{eq:cmmr_effective_energy}.

The geodesic equations, analogous to Eqs.~\eqref{eq:ode_for_rho_motion} and \eqref{eq:ode_for_z_motion}, can be derived from the Euler-Lagrange equations for the Lagrangian \eqref{eq:cmmr_lagrangian}. Their explicit forms, in terms of $E$, $L$ and the metric functions $f$, $\omega$ and $e^{2\gamma}$, are given in Eqs.~(16) and (17) of Ref.~\cite{Dubeibe2016}. Once $\rho(\lambda)$ and $z(\lambda)$ are known, $t(\lambda)$ and $\phi(\lambda)$ are determined by direct integration of Eqs.~\eqref{eq:cmmr_energy_t_phi} and \eqref{eq:cmmr_ang_mom_t_phi}.

\subsection{Ergosphere}

Since the CMMR metric is stationary and we are considering neutral particles in geodesic motion, we use the standard definition of an ergosphere to study the possibility of negative energy orbits and energy extraction. In other words, the ergosphere is the region where the time translation Killing vector field becomes spacelike, i.e.~$(\partial_t)^\mu (\partial_t)_\mu > 0$. Taking into account the line element \eqref{eq:cmmr_line_element}, it is straightforward to show that the ergosphere of the CMMR spacetime is the locus of points that satisfy
\begin{equation}
    f(\rho,z) < 0.
    \label{eq:cmmr_ergo_ineq}
\end{equation}

\begin{figure}[!htbp]
\centering
  \includegraphics[width = 0.95 \linewidth]{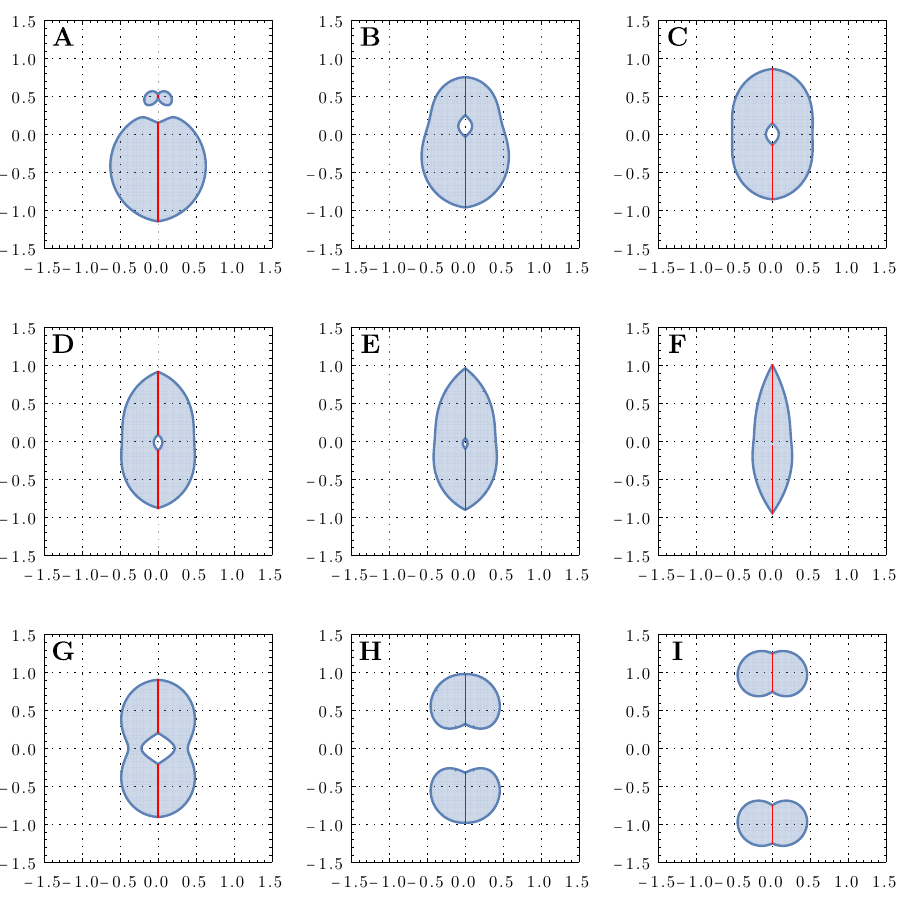}
  \caption{The $\phi=0$ section of the ergosphere of the CMMR metric for different set of parameters labeled \textbf{A}-\textbf{I} (see Table  \ref{tab:cmmr_ergo_tab}). In each plot the horizontal and vertical axes are $\rho/M_T$ and $z/M_T$, respectively. The red lines indicate the location of the black hole's horizons.}
  \label{fig:cmmr_ergospheres} 
\end{figure}

We sketch this ergosphere in Fig.~\ref{fig:cmmr_ergospheres}, where each panel is labeled by a letter (\textbf{A}-\textbf{I}) and corresponds to a different set of parameters (which are specified in Table \ref{tab:cmmr_ergo_tab}). In each panel, the blue shaded region represents the $\phi = 0$ section of the ergosphere, while the red lines represent the event horizons of the black holes. The top row of the figure (panels \textbf{A}-\textbf{C}) exhibits the effect of changing the mass ratio of the system while keeping the total mass, both spins and the separation parameter fixed. It shows that, analogously to the MP case, initially disjoint ergospheres may merge into a single connected ergosphere when the mass ratio increases. The middle row of Fig.~\ref{fig:cmmr_ergospheres} (panels \textbf{D}-\textbf{E}), on the other hand, shows the effect of changing the spin parameter of the top black hole while keeping all other parameters fixed. We observe that when initially aligned spins become anti-aligned, the ergosphere becomes thinner and elongated along the symmetry axis. Finally, the bottom row (panels \textbf{G}-\textbf{H}) illustrates the effect of increasing the separation parameter when all other parameters are kept fixed. Similarly to what happens in the MP case, if the distance between the black holes is sufficiently large, there will be two disconnected ergospheres, one for each black hole. 

\begin{table}[!htbp]
    \begin{ruledtabular}
    \begin{tabular}{ccccc}
        Panel      & $M_1/M_2$    & $a_1/M_T$   & $a_2/M_T$  & $R/M_T$    \\
        \textbf{A} & $0.16$ & $0.65$  & $0.65$ & $1.00$  \\
        \textbf{B} & $0.58$ & $0.65$  & $0.65$ & $1.00$  \\
        \textbf{C} & $1.00$  & $0.65$  & $0.65$ & $1.00$  \\
        \textbf{D} & $1.00$  & $0.50$  & $0.65$ & $1.00$  \\
        \textbf{E} & $1.00$  & $0.30$  & $0.65$ & $1.00$  \\
        \textbf{F} & $1.00$  & $-0.10$ & $0.65$ & $1.00$  \\
        \textbf{G} & $1.00$  & $0.65$  & $0.65$ & $1.11$ \\
        \textbf{H} & $1.00$  & $0.65$  & $0.65$ & $1.30$ \\
        \textbf{I} & $1.00$  & $0.65$  & $0.65$ & $2.00$  \\
    \end{tabular}
    \end{ruledtabular}
    \caption{Parameters that define the CMMR metrics associated with the ergospheres \textbf{A}-\textbf{I} shown in Fig.~\ref{fig:cmmr_ergospheres}.}
    \label{tab:cmmr_ergo_tab}
\end{table}

\subsection{Bound negative energy orbits and the Penrose Process}

\begin{figure}[!htbp]
\centering
  \includegraphics[width = 0.95 \linewidth]{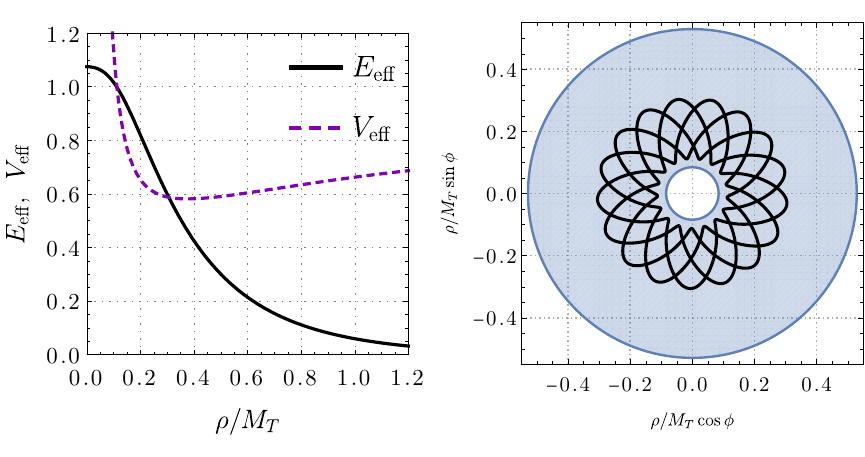}
    \caption{Left panel: effective energy (black curve) and effective potential (purple dashed curve) for $L=-2.5$ and $E=-0.054$, when the CMMR metric is characterized by $a_1=a_2=0.65$, $M_1 = M_2 = 0.5$, and $R = 1$ (corresponding to the label \textbf{C} in Table \ref{tab:cmmr_ergo_tab}). The turning points are located at $\rho_{-}/M_T = 0.112531$ and at $\rho_{+}/M_T = 0.306081$. Right panel: the associated trajectory in the $z = 0$ plane when $\rho(0)/M_T=0.25$ and $\phi(0)=0$. The blue region is the $z=0$ section of the ergosphere of the spacetime.}
  \label{fig:cmmr_veff} 
\end{figure}

To demonstrate the existence of bound negative energy orbits and the possibility of using them to extract energy from non-coalescing Kerr binaries, we shall restrict our attention to systems of equal mass and spin. This symmetry allows for the existence of stable orbits (in the sense already discussed for the MP metric) in the $z=0$ plane. To find a negative energy trajectory that is confined outside the black holes, we choose the energy $E$ and the angular momentum $L$ such that there are two orbital turning points of Eq.~\eqref{eq:effective1} that lie inside the ergosphere of the system. Similarly to what was done in the MP case, once the initial radius $\rho(0)$, the energy, and the angular momentum are fixed, we solve Eq.~\eqref{eq:cmmr_energy_rho_z} to determine $\dot{\rho}(0)$ and integrate the geodesic equations. Using the parameters that produce the ergosphere \textbf{C} of Fig.~\ref{fig:cmmr_ergospheres} and Table \ref{tab:cmmr_ergo_tab}, we show an example of such a negative energy orbit in Fig.~\ref{fig:cmmr_veff} (right panel). The corresponding effective potential and effective energy are also shown in Fig.~\ref{fig:cmmr_veff} (left panel).

\begin{figure}[!htbp]
\centering
  \includegraphics[width = 0.95 \linewidth]{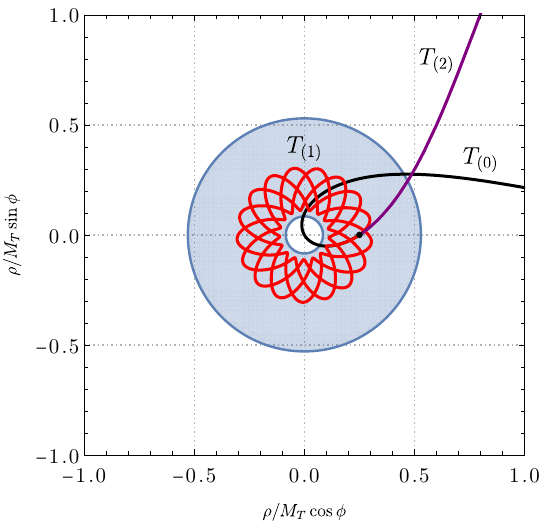}
  \caption{Penrose process in the $z=0$ plane of a CMMR spacetime with $M_1=M_2=0.5$, $a_1=a_2=0.65$, and $R=1$. The incoming trajectory $T_{(0)}$ (black curve) splits at the black point ($\rho/M_T=0.25$, $\phi=0$) into the negative energy orbit $T_{(1)}$ (red curve) and the trajectory of the escaping fragment $T_{(2)}$ (purple curve). The parameters that generate these trajectories are shown in Table \ref{tab:cmmr_penrose_example}. The blue region is the $z=0$ section of the ergosphere (for particle 1).}
 \label{fig:cmmr_penrose} 
\end{figure}

Adopting the same notation introduced in Sec.~III for the trajectories in a Penrose process around the MP black hole, and taking advantage of the negative energy orbit depicted in Fig.~\ref{fig:cmmr_veff}, we now consider the possibility of energy extraction in the CMMR spacetime. By employing the conservation of 4-momentum (as in Sec.~III), we construct an explicit example of a Penrose process. The obtained trajectories are shown in Fig.~\ref{fig:cmmr_penrose} and the corresponding parameters are given in Table \ref{tab:cmmr_penrose_example}. The efficiency of the process, calculated through Eq.~\eqref{eq:penrose_efficiency_general}, is $\eta \approx 0.08 \%$.

\renewcommand{\arraystretch}{1.2}
\begin{table}[!htbp]
\begin{ruledtabular}
\begin{tabular}{lllllll}
$i$ & $m_{(i)}/m_0$ & $E_{(i)}$  & $L_{(i)}$  & $\dot{\rho}_{(i)}$ & $\dot{z}_{(i)}$ \\ \vspace{-0.3cm} \\
0   & 1.0000000             & 2.00000         & 0         .000000 & 4.343904           & 0               \\
1   & 0.0289697     & -0.05400     & -2.500000       & 0.313887           & 0               \\
2   & 0.3148980     & 6.35623    & 0.229993   & 13.765800            & 0               \\
\end{tabular}
\end{ruledtabular}
\caption{Parameters that generate the trajectories $T_{(0)}$, $T_{(1)}$, and $T_{(2)}$ of the Penrose process shown in Fig.~\ref{fig:cmmr_penrose}. The derivatives $\dot{\rho}_{(i)}$ and $\dot{z}_{(i)}$ are evaluated at the break-up point.}
\label{tab:cmmr_penrose_example}
\end{table}

\section{Final remarks}

We have demonstrated the possibility of energy extraction by the Penrose process in a static binary black hole described by the MP spacetime. Relying on the concept of a particle dependent generalized ergosphere, we have shown that closed orbits of negative energy exist outside of the event horizons of a MP binary and can be used in energy extraction processes. We have also studied the efficiency of the Penrose processes in MP black holes, understanding how it is influenced by several parameters and providing a prescription for maximizing the extraction of energy. Even though the details of realistic processes around astrophysical binary black holes are much more complicated, we were able to illustrate that some of the conclusions reached for the MP spacetime do generalize for a spacetime consisting of two Kerr black holes separated by a ``strut".

Thanks to the success of the LIGO-Virgo collaboration, we live in an age where direct gravitational wave detections provide evidence on the abundance of compact binaries in our Universe. For example, in LIGO's latest observation run, more than $90$ binary system candidates were identified~\cite{LIGO_CATALOG}.
Nevertheless, the possibility of energy extraction from such systems by the Penrose process is yet to be explored. In view of that, the MP and the CMMR metric can give a fair idea of what happens in a ``snapshot'' of the collision process between two black holes and can be considered a first step towards a better understanding of effects such as the Penrose process in astrophysically relevant setups. And a better understanding of the Penrose process in black hole binaries may also contribute to a better understanding of superradiant effects~\cite{Misner,zeldovich1,staro1,Bekenstein:1973mi,Brito:2015oca} in similar scenarios. In fact, even though some interesting ideas regarding superradiance by binary systems have been proposed~\cite{Rosa:2015hoa,Rosa:2016bli,Wong:2019kru}, further research is needed.   


\acknowledgments

This research was partially financed by the Coordena\c{c}\~ao de Aperfei\c{c}oamento de Pessoal de N\'{i}vel Superior (CAPES, Brazil) - Finance Code 001. M.~R.~acknowledges support from the Conselho Nacional de Desenvolvimento Cient\'{i}fico e Tecnol\'{o}gico (CNPq, Brazil), Grant No. FA 315664/2020-7.

\bibliographystyle{apsrev4-2}
\bibliography{ref}

\end{document}